\documentclass[12pt,preprint]{aastex}

\begin{document}
\slugcomment{ApJ submitted 2009.01.15, accepted 2009.11.03}
\title{Dust Transport in Protostellar Disks\linebreak
  Through Turbulence and Settling}

\author{N.~J.~Turner\altaffilmark{1}, A.~Carballido\altaffilmark{1,2}
  and T.~Sano\altaffilmark{3}}
\altaffiltext{1}{Jet Propulsion Laboratory, California Institute of
  Technology, Pasadena, California 91109, USA;
  neal.turner@jpl.nasa.gov}
\altaffiltext{2}{Instituto de Astronom\'ia, Universidad Nacional
  Aut\'onoma de M\'exico, DF 04510, M\'exico}
\altaffiltext{3}{Institute of Laser Engineering, Osaka
  University, Suita, Osaka 565-0871, Japan}

\begin{abstract}
  We apply ionization balance and MHD calculations to investigate
  whether magnetic activity moderated by recombination on dust grains
  can account for the mass accretion rates and the mid-infrared
  spectra and variability of protostellar disks.  The MHD calculations
  use the stratified shearing-box approach and include grain settling
  and the feedback from the changing dust abundance on the resistivity
  of the gas.  The two-decade spread in accretion rates among
  Solar-mass T~Tauri stars is too large to result solely from
  variations in the grain size and stellar X-ray luminosity, but can
  plausibly be produced by varying these parameters together with the
  disk magnetic flux.  The diverse shapes and strengths of the
  mid-infrared silicate bands can come from the coupling of grain
  settling to the distribution of the magneto-rotational turbulence,
  through the following three effects.  First, recombination on grains
  1~$\mu$m or smaller yields a magnetically-inactive dead zone
  extending more than two scale heights from the midplane, while
  turbulent motions in the magnetically-active disk atmosphere
  overshoot the dead zone boundary by only about one scale height.
  Second, grains deep in the dead zone oscillate vertically in wave
  motions driven by the turbulent layer above, but on average settle
  at the rates found in laminar flow, so that the interior of the dead
  zone is a particle sink and the disk atmosphere will become
  dust-depleted unless resupplied from elsewhere.  Third, with
  sufficient depletion, the dead zone is thinner and mixing dredges
  grains off the midplane.  The last of these processes enables
  evolutionary signatures such as the degree of settling to sometimes
  decrease with age.  The MHD results also show that the magnetic
  activity intermittently lifts clouds of small grains into the
  atmosphere.  Consequently the photosphere height changes by up to
  one-third over timescales of a few orbits, while the extinction
  along lines of sight grazing the disk surface varies by factors of
  two over times down to a tenth of an orbit.  We suggest that the
  changing shadows cast by the dust clouds on the outer disk are a
  cause of the daily to monthly mid-infrared variability found in some
  young stars.
\end{abstract}


\section{INTRODUCTION\label{sec:intro}}

Concentration of the primordial solid material is a fundamental
requirement for planet formation.  The interstellar medium contains
about 1\% solids by mass in the form of dust grains, and the Sun has a
similar abundance of heavy elements, while the terrestrial planets
consist almost entirely of refractory material.  Even the most
gas-rich Solar system planet, Jupiter, has between 1.5 and 6~times the
Solar abundance of heavy elements \citep{sg04}.

A basic concentration process operating in the Solar nebula and other
protostellar disks is the settling of dust particles due to the
component of the stellar gravity perpendicular to the disk plane.  The
grains settle at different speeds depending on their ratio of mass to
area, leading to mutual collisions.  If the disk gas is laminar and
the initially sub-micron particles readily stick together, then most
of the solid material settles into a thin midplane layer within a few
tens of thousands of years at 1 to 10~AU \citep{w80,nn81,ns86}.  Such
rapid loss of the grains conflicts with observations of scattered
starlight showing some dust remains suspended in the atmospheres of
million-year-old protostellar disks \citep{bs96,sm03,ws04}.  However
mid-infrared colors suggest the dust abundance is reduced near the
disk photosphere, consistent with the incorporation of some material
into larger bodies in the interior \citep{fh06}.  The smaller
particles can remain aloft if stirred by turbulence in the gas
\citep{dd04b}.

Turbulence is also commonly invoked to drive the observed flow of
material onto the central star.  In particular, the turbulence
resulting from the magneto-rotational instability
\citep[MRI;][]{bh91,bh98} can provide the necessary outward transfer
of orbital angular momentum.  However, the MRI requires a sufficient
level of ionization for coupling the disk gas to the magnetic fields.
The weak internal heating means thermal ionization is effective only
very near the star, while the interstellar cosmic rays and stellar
X-rays ionize only the top and bottom surfaces of the disk, failing to
penetrate the interior \citep{g96,ig99}.  Furthermore, the
recombination of the free charges on grain surfaces is so efficient
that a small mass fraction of sub-micron particles is sufficient to
shut off MRI turbulence in much of the region where the planets formed
\citep{sm00}.  As a result, protostellar disks consist of a laminar
dead zone sandwiched between two turbulent active layers.  The dead
zone extends from 0.1 to 15~AU in the minimum-mass Solar nebula, given
micron-sized grains and ionization by the median stellar X-ray
luminosity \citep{td09}.

This paper deals with the consequences of the active and dead layers
for dust settling and disk evolution.  We investigate how the gas
flows govern the distribution of the grains, and how the grains in
turn affect the dead zone size, using the methods outlined in
\S\ref{sec:methods} to make ionization balance and 3-D MHD
calculations of small patches of the disk around a T~Tauri star.
Since a wealth of data is now available on the abundance, size and
composition of dust in the surface layers of protostellar disks
overlying the regions where planets are likely to form
\citep{fh06,sw06,bh08,wl09}, we examine the observational signatures
of the movements of the grains, seeking to understand the following
issues.
\begin{enumerate}
\item What is the cause of the large measured range in accretion rate
  at a given stellar mass (\S\ref{sec:massflow})?  Solar-mass T~Tauri
  stars grow at rates between about $10^{-9}$ and $10^{-7}$~M$_\odot$
  \citep[compiled by][]{ml05}, while simple magnetic accretion models
  have a fixed column of accreting gas set by the penetration depth of
  the ionizing radiation, implying a unique mass flow rate
  \citep{hd06}.
\item Why do T~Tauri stars show such diverse mid-infrared spectra
  (\S\ref{sec:movements})?  The variety in the silicate band strengths
  and shapes indicates wide ranges in the abundances of small and
  crystalline particles.  High crystalline mass fractions appear more
  often in systems having colors consistent with grains concentrated
  near the midplane \citep{fh06,wl09}.
\item What are the origins of the variability over intervals of days
  to years in the 10-$\mu$m silicate feature \citep{wh00,ww04,bl09}
  and nearby continuum \citep{la08,mf09} (\S\ref{sec:variable})?  The
  shorter variation timescales are substantially faster than the
  orbital periods at the AU radii where the feature forms.
\end{enumerate}
A summary, discussion and conclusions are in \S\ref{sec:conc}.

\section{METHODS\label{sec:methods}}

We consider a patch of disk lying 5~AU from a Solar-mass star, where
the orbital period is 11.2~yr.  The surface density $\Sigma=152$
g~cm$^{-2}$ and temperature $T=125$~K are taken from the minimum-mass
Solar nebula \citep{hn85}.  We solve the standard equations of
isothermal resistive MHD \citep{fs03} in the frame co-rotating with
the orbital speed of the domain center, using the stratified
shearing-box approximation \citep{hg95,ms00} with the ZEUS code
\citep{sn92a,sn92b}.  The density in the initial hydrostatic
equilibrium varies with the height $z$ as a Gaussian
$\rho=(\Sigma/\sqrt{2\pi}H)\exp(-z^2/2H^2)$ having scale height
$H=c_s/\Omega=0.25$~AU equal to the ratio of the sound speed $c_s$
with the orbital frequency $\Omega$.  The mean molecular weight of the
gas is $2.3$.  The domain size along the local Cartesian axes $(x, y,
z)$, corresponding to the radial, azimuthal and vertical directions,
is $L_x\times L_y\times L_z = 0.5\times 2\times 2$~AU and the box is
divided into $32\times 64\times 128$~grid cells.  The $x$-boundaries
are shearing-periodic, the $y$-boundaries are periodic and the
$z$-boundaries allow outflow but no inflow.

A major factor in fixing the strength of magneto-rotational turbulence
is the net vertical flux of the magnetic field \citep{hg95,si04,pc07}.
The periodic horizontal boundaries in our calculations mean that the
initial vertical flux is preserved over time, while any
spatially-varying fields are readily destroyed through reconnection.
We choose a starting field that is the sum of a uniform vertical part
$B_1$ and a zero-net-flux part with strength $B_2$, having a
sinusoidal variation in $x$ with one period in the box width.  The
initial magnetic field thus has components $B_x=0$, $B_y=B_2\cos 2\pi
x/L_x$ and $B_z=B_1+B_2\sin 2\pi x/L_x$.  The uniform part is chosen
on the basis of the solar nebula fields of 0.1 to several Gauss,
recorded in the remanent magnetization of meteorites from the asteroid
belt \citep{ch91}.  The fields at 5~AU were more likely near the low
end of this range.  Magneto-rotational turbulence typically produces
fields with an RMS vertical component that is 10 to 20\% of the total
field strength \citep[e.g.][]{ms00}.  From these considerations, we
choose $B_1=6$~mG.  The spatially-varying part of the initial magnetic
field is given an amplitude $B_2=200$~mG similar to the RMS field
strength found in ideal-MHD test calculations.  The pressure in the
uniform part of the field is $51\,000$ times less than the initial gas
pressure at the midplane, and $18$~times less than the gas pressure at
four scale heights.

The resistivity in each grid cell is computed by solving a
time-dependent ionization network including recombination in the gas
phase and on grains (\S\ref{sec:resistivity}).  We carry out three
resistive MHD calculations, each involving grains of one size, either
1, 10 or 100~$\mu$m, modeled as a fluid settling through the gas at
the terminal speed (\S\ref{sec:laminar}).  Initially the grains are
well-mixed with a mass fraction of 1\% and the reaction network is in
chemical equilibrium.  The largest grains settle two orders of
magnitude faster than the smallest.  In addition their cross section
per unit gas mass is two orders of magnitude less, producing slower
recombination and a smaller dead zone.

\subsection{Coupling of the Gas to Magnetic Fields\label{sec:resistivity}}

The electrical resistivity of the disk gas
\begin{equation}\label{eq:resistivity}
\eta=234\,{\sqrt{T}\over x_e}\,{\rm cm^2 s^{-1}}
\end{equation}
is inversely proportional to the electron fraction $x_e=n_e/n_n$,
where $n_e$ is the electron number density and $n_n$ the total number
density of neutrals \citep{bb94}.  We compute the electron fraction by
solving the reduced gas-phase recombination reaction network of
\cite{od74}, augmented with reactions involving grains as described by
\cite{in06a}.  The electrons recombine with the ions through grain
charging and neutralization, and in the gas phase through dissociative
recombination and through charge transfer to metal atoms followed by
radiative recombination.  The fraction of the metal atoms free to
leave the grain surfaces and enter the gas phase is chosen to be 1\%.
The advection of the reacting species across the grid is treated in an
operator-split fashion \citep{ts07}.

\begin{figure}[tb!]
  \epsscale{0.5}
  \plotone{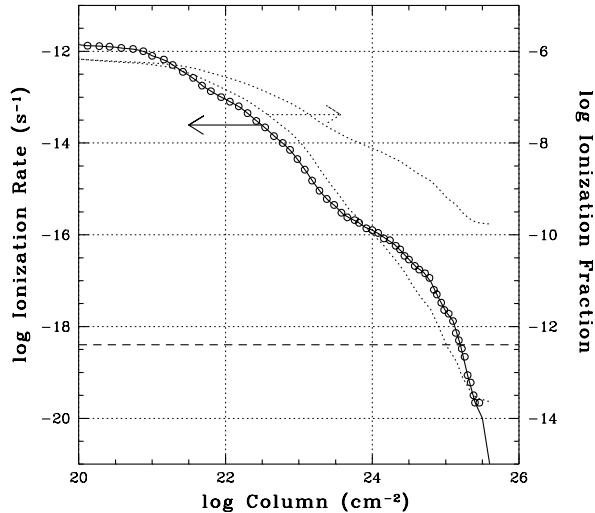}

  \figcaption{\sf Ionization rate and electron fraction versus column
    at 5~AU in the minimum-mass Solar nebula.  Circles show X-ray
    ionization rates from \cite{ig99} on the left-hand scale as
    indicated by the solid arrow, with the solid line marking the
    extrapolation to greater column depth.  The stellar X-ray
    luminosity is $2\times 10^{30}$~erg~s$^{-1}$.  A horizontal dashed
    line indicates the $^{26}$Al decay ionization rate for well-mixed
    dust.  The dotted curves show the initial ionization fraction with
    no dust (upper) and with 1-$\mu$m grains (lower), on the
    right-hand scale as indicated by the dotted arrow.  The decline in
    ionization fraction with depth is closely tied to the attenuation
    of the X-rays.
  \label{fig:ionization}}
\end{figure}

The ionization is due to the X-rays from the young star and the
radioactive decay of $^{26}$Al nuclei within the disk.  The X-ray
ionization rate is interpolated in Monte Carlo radiative transfer
results including scattering, computed by \cite{ig99}.  At depths not
shown on their plot we attenuate the X-rays using an $e$-folding
column of 8~g~cm$^{-2}$.  The X-rays have a 5~keV thermal spectrum.
Where not otherwise specified we use a fiducial stellar X-ray
luminosity of $2\times 10^{30}$~erg~s$^{-1}$, matching young
Solar-mass stars observed in the Orion nebula \citep{gf00}.  The
ionization rate is proportional to the luminosity.  The $^{26}$Al
radioactive decay ionization rate is $4\times 10^{-19}$~s$^{-1}$ at
the initial dust-to-gas mass ratio of 0.01, and varies in proportion
to the local dust abundance.  The X-rays dominate the ionization in
the outermost 30~g~cm$^{-2}$, or the top and bottom 20\% of the mass
column (Figure~\ref{fig:ionization}).  We neglect the changes in the
X-ray ionization rate caused by the variation of the opacity with the
dust abundance \citep{gn97}, since no detailed radiative transfer
calculations are available.  In line with the absence of heavy-element
opacities in the Monte Carlo calculations, we also neglect the
absorption of X-rays in the grains.

\subsection{Treatment of Grain Settling\label{sec:laminar}}

\begin{figure}[tb!]
  \epsscale{0.5}
  \plotone{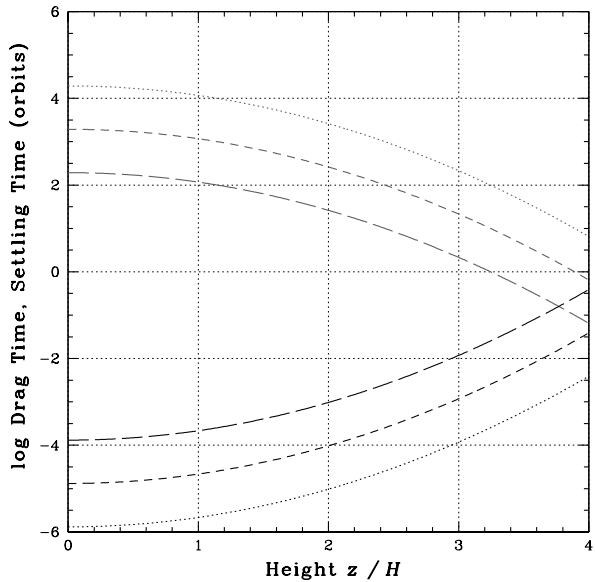}

  \figcaption{\sf Stopping time due to gas drag (lower black curves)
    and settling timescale (upper gray curves) versus height at 5~AU
    in the minimum-mass Solar nebula, for compact spherical particles
    with radii 1 (dotted), 10 (dashed) and 100~$\mu$m (long-dashed
    curve).  The settling times are from eq.~\ref{eq:tsett}.
  \label{fig:tstop}}
\end{figure}

The drag force between the gas and dust follows the Epstein law, since
the mean free path of the molecules in the gas is much larger than the
grain size, and the particles generally move through the gas at
subsonic speeds.  The acceleration felt by dust particles of radius
$a$ and internal density $\rho_d$ moving at speed $v$ through gas with
density $\rho$ and sound speed $c_s$ is
\begin{equation}
f=\left(\rho\over\rho_d\right)\left(c_s\over a\right) v.
\end{equation}
The particles come to rest if no other forces act after the
drag time
\begin{equation}\label{eq:tdrag}
t_D=v/f=\left(\rho_d\over \rho\right)\left(a\over c_s\right).
\end{equation}
Figure~\ref{fig:tstop} shows the variation of the drag time with
height and grain size in the hydrostatic equilibrium density profile.
For grains 100~$\mu$m and smaller, the drag time is less than the
orbital period within four scale heights of the midplane.  We have
chosen an internal density for the particles $\rho_d=5$~g~cm$^{-3}$
about twice that of rock, so that the grains will settle significantly
within the 60-orbit time span of the MHD calculations.  The geometric
cross-section per unit gas mass, a key to determining the
recombination rate, is thus equal to that of particles with a more
typical internal density $2.5$~g~cm$^{-3}$ and a lower dust-to-gas
mass ratio 0.5\%.

Particles falling through the gas under gravity after one drag time
reach a steady-state in which the two forces are equal and opposite.
Balancing the drag force $v/t_D$ with the vertical acceleration due to
gravity in the disk, $\Omega^2 z$, gives the terminal speed
\begin{equation}\label{eq:vterminal}
v_T=\Omega^2 z t_D.
\end{equation}
This amounts to $\Omega t_D$ times the sound speed for each scale
height of distance from the midplane.  The time to settle to a height
$z$ is approximately
\begin{equation}\label{eq:tsett}
t_S=z/v_T=(\Omega^2 t_D)^{-1}
\end{equation}
or, in words, the settling time in orbits is $(2\pi)^{-2}$ or about
one-fortieth the inverse of the drag time in orbits.

\begin{figure}[tb!]
  \epsscale{0.5}
  \plotone{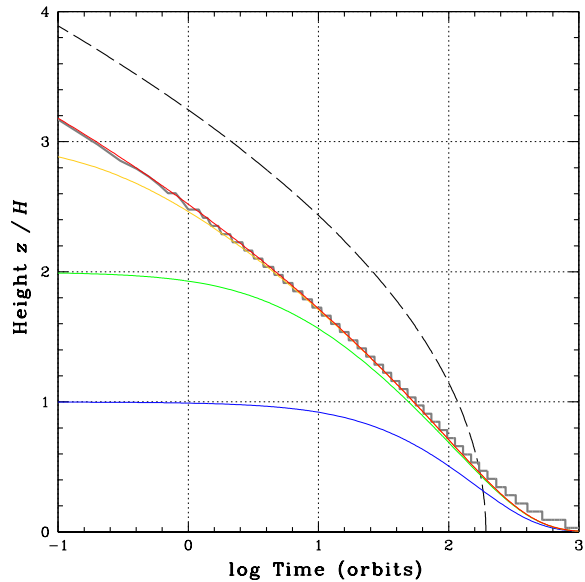}

  \figcaption{\sf Dust layer thickness versus time at 5~AU in a
    laminar, minimum-mass Solar nebula.  The grains are compact
    spheres 100~$\mu$m in radius.  The dust is initially well-mixed
    and settles as a fluid in the ZEUS code (heavy gray line with
    steps corresponding to the grid spacing) and as individual
    particles followed with a Runge-Kutta trajectory integrator
    (colored curves) starting at 1, 2, 3 and 4~scale heights.  The
    dashed line shows the rough estimate from eq.~\ref{eq:tsett}.
  \label{fig:zsett}}
\end{figure}

We wish to model the grain settling in magneto-rotational turbulence.
Since the drag time is less than the correlation time $t_C\approx
1/\Omega$ of the gas motions in the turbulence \citep{fp06,tw06}, the
particles settle at their terminal speeds almost always.  An exception
occurs for 100-$\mu$m particles near $4H$, which have drag times
comparable to the correlation time, but as shown by the estimates
above and confirmed in \S\ref{sec:movements}, these particles quickly
settle into the interior.  We treat the dust as a fluid moving with a
velocity equal to the sum of the gas velocity ${\bf v}$ and the
terminal velocity ${\bf v}_T$, with the latter directed toward the
midplane.  When the gas is near hydrostatic equilibrium, this is
equivalent to the short friction time approximation used by
\cite{jk05}, and has the advantage that the time step can be longer
than the drag time, reducing the computing load by several orders of
magnitude compared with the full dust equation of motion.  The
back-reaction of the dust on the gas is neglected.  We solve the
conservation equation for the grain number density $n$,
\begin{equation}
  {\partial n\over\partial t} + {\bf\nabla}\cdot n({\bf v}+{\bf v}_T)=0,
\end{equation}
using the same standard second-order accurate van Leer transport
algorithm applied to the gas.  The Courant condition for the dust is
satisfied in our MHD calculations because the terminal speed is less
than the greatest Alfv\'en speed on the grid, typically about $13c_s$.
To test the method, we compute the time evolution of a layer of
100~$\mu$m dust particles initially well-mixed with the gas.  The
domain extends from the midplane to four scale heights and is divided
into 64~grid cells.  The gas is stationary while the dusty layer
contracts over time through settling.  To gauge the accuracy of the
solution, we also compute the paths followed by individual grains
using a fourth-order Runge-Kutta integration scheme \citep{tw06}.  The
results are shown in Figure~\ref{fig:zsett}.  The settling is fastest
in the low-density disk atmosphere, so that most of the grains that
have settled significantly at any given time $t$ lie bunched-up near
the top of the dusty layer.  The thickness of the layer varies roughly
as the square root of the logarithm of $1/t$, in agreement with
eq.~\ref{eq:tsett} given the Gaussian gas density profile.
Equation~\ref{eq:tsett} overestimates the thickness at early times,
when the distance the particles have fallen is less than their height,
and underestimates the thickness after a few hundred years, when the
edge of the dust layer lies near the midplane, the gas density is
roughly uniform, and the terminal speed no longer decreases sharply as
the particles fall.  The paths of four individual grains from the
trajectory integrator starting at 1, 2, 3 and 4~scale heights
illustrate the pile-up near the top of the dusty layer and show that
the treatment of the settling in ZEUS has satisfactory accuracy.  The
uppermost particle (red curve) lies near to the highest grid cell in
the ZEUS calculation having a grain abundance more than half the
initial value (gray line).  After a few hundred years, when the dusty
layer is just a few cells thick, the ZEUS calculation is less accurate
as numerical diffusion becomes important and spreads the layer.
Similar results are obtained in test calculations with 10- and
1-$\mu$m grains where settling is 10 and 100 times slower.

\section{MASS FLOW RATE
  \label{sec:massflow}}

Conservation of orbital angular momentum dictates that the rate at
which mass flows through the disk to the star
\begin{equation}
{\dot M} = -{2\pi\over\Omega}\int w_{xy}\,dz
\end{equation}
is proportional to the height-integrated accretion stress $w_{xy}$
when winds are negligible.  Since most of the mass flow passes through
the active layers, the accretion rate is the product of the active
layer thickness with the volume-averaged stress there.

The thickness of the active layer depends on the ionization profile of
the disk material.  Magneto-rotational turbulence can be shut off by
the Ohmic resistivity when charged and neutral particles collide
frequently \citep{j96} and by ambipolar diffusion when collisions are
rare \citep{bb94,hs98}.  The third non-ideal term in the induction
equation, the Hall term, by contrast has little effect on the
saturation amplitude of the turbulence over the range of parameters
that has been explored in non-linear calculations \citep{ss02b}.  The
active layer thickness also depends on the field strength.  The linear
magneto-rotational instability grows slowly if the magnetic pressure
exceeds the gas pressure \citep{ko00}, on fields having the
substantial toroidal component expected given the rapid orbital shear.
Numerical ideal-MHD calculations, in general agreement with the linear
results, show the stress declines with height in the
strongly-magnetized disk corona \citep{ms00}.

\begin{figure}[tb!]
  \epsscale{0.5}
  \plotone{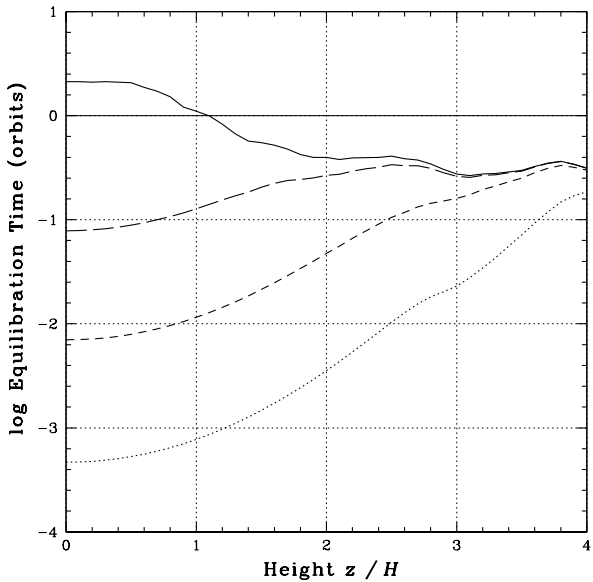}

  \figcaption{\sf Ionization equilibration timescale versus height at
    5~AU in the minimum-mass solar nebula.  The gas contains
    well-mixed grains with radii 1 (dotted), 10 (dashed) and
    100~$\mu$m (long-dashed curves).  A solid line indicates the case
    with grains absent.  \label{fig:teqm}}
\end{figure}

To estimate the active layer thickness, we compare the limits imposed
by the Ohmic and ambipolar terms and the magnetic pressure.  We bring
the ionization-recombination reaction network to local equilibrium in
the presence of well-mixed grains.  The electron density approaches
within 10\% of its equilibrium value in less than one orbit, at all
locations below 4~scale heights when grains are present
(Figure~\ref{fig:teqm}).  Consequently, turbulent mixing occurring on
timescales longer than the orbital period is unlikely to change the
ionization significantly.

\begin{figure}[tb!]
  \epsscale{0.5}
  \plotone{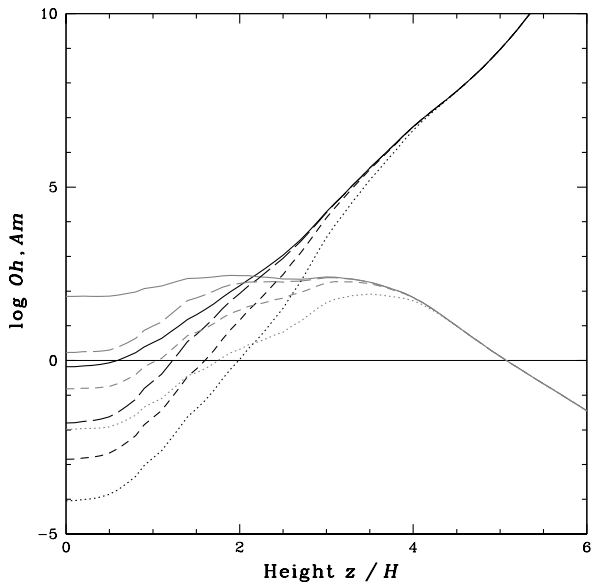}

  \figcaption{\sf Dimensionless numbers measuring the importance of
    Ohmic dissipation (black curves) and ambipolar diffusion (gray
    curves) relative to induction in the magnetic field evolution
    equation, as functions of height at 5~AU in the minimum-mass solar
    nebula.  A horizontal line marks the critical value of unity above
    which the magneto-rotational instability can drive turbulence.
    The gas contains well-mixed grains with radii 1 (dotted), 10
    (dashed) and 100~$\mu$m (long-dashed curves).  Solid lines
    indicate the case with grains absent.  The strength of the
    vertical component of the magnetic field is 6~mG.  The extent of
    the magneto-rotational turbulence is limited by the Ohmic
    resistivity near the midplane, and by the ambipolar diffusion
    above $5H$.  \label{fig:dimensionless}}
\end{figure}

The active layer separates the dead zone inside the disk from the
corona outside.  Rapid Ohmic dissipation prevents magneto-rotational
turbulence at the high densities found in the dead zone, and rapid
ambipolar diffusion prevents turbulence at the lower densities found
in the corona \citep{w07}.  The minimum Ohmic and ambipolar diffusion
rates needed to shut off the turbulence correspond to the field
drifting across the characteristic vertical wavelength of the
instability, $v_{Az}/\Omega$, within the characteristic growth time
$1/\Omega$, where the Alfv\'en speed $v_{Az}=B_z/\sqrt{4\pi\rho}$.
The resulting Ohmic dead zone criterion is
\begin{equation}\label{eq:elsasser}
  Oh \equiv {v_{Az}^2 \over \eta\Omega} < 1
\end{equation}
\citep{si01,ss02b}.  The Elsasser number criterion
eq.~\ref{eq:elsasser} yields more accurate estimates of the dead zone
boundary than a version with the Alfv\'en speed replaced by the sound
speed \citep{ts07}.  The dimensionless number governing whether
ambipolar diffusion prevents turbulence in the neutrals
\begin{equation}\label{eq:ambipolar}
  Am \equiv \gamma\rho_i/\Omega < 1
\end{equation}
\citep{hs98,ss02a,cm07} is proportional to the ion-neutral drag
coefficient $\gamma$ and the ion density $\rho_i$, and independent of
the magnetic field strength.  Using the drag coefficient described by
\cite{ss02a}, we find that eq.~\ref{eq:ambipolar} holds above about
$5H$ (Figure~\ref{fig:dimensionless}).  The height where $Am$ reaches
unity is almost independent of the grain properties, because most
recombination occurs in the gas phase at the low ambient densities and
high ionization fractions \citep{w07}.

The height where the magnetic and gas pressures match depends on the
strength of the magnetic fields generated in the turbulence.  In our
three resistive MHD calculations, the sinusoidal part of the magnetic
field quickly dissipates throughout, leaving in the dead zone just the
net vertical part $B_1$, and in the active layers the vertical part
plus a newly-regenerated tangled field.  The horizontally-averaged
Maxwell stress outside the dead zone is a few times the Reynolds
stress, as expected in MRI turbulence.  The magnetic pressure averaged
over the time interval from 10 to 60~orbits exceeds the gas pressure
above about $2.7H$, and the stress declines steeply with height near
the domain boundary at $4H$.  Ideal-MHD calculations in a taller
shearing-box \citep{ms00} show an exponential decline in the stress
through $7c_s/\Omega$.  These results indicate that the upper boundary
of the active layer is determined by the strong magnetization, rather
than the ambipolar diffusion, and lies near $4H$.

At the base of the active layer the Ohmic resistivity given by
eq.~\ref{eq:resistivity} shuts off the turbulence according to
eq.~\ref{eq:elsasser}.  Combining these relations under the conditions
found at 5~AU in the minimum-mass Solar nebula yields a critical
electron fraction $x_e=5\times 10^{-15}\beta_z$.  The ratio $\beta_z$
of the gas pressure to the mean pressure in the vertical component of
the field is about $10^4$ near the base of the active layer in our
three MHD calculations, so the ionization fraction is a few times
$10^{-11}$.

\begin{figure}[tb!]
  \epsscale{1}
  \plottwo{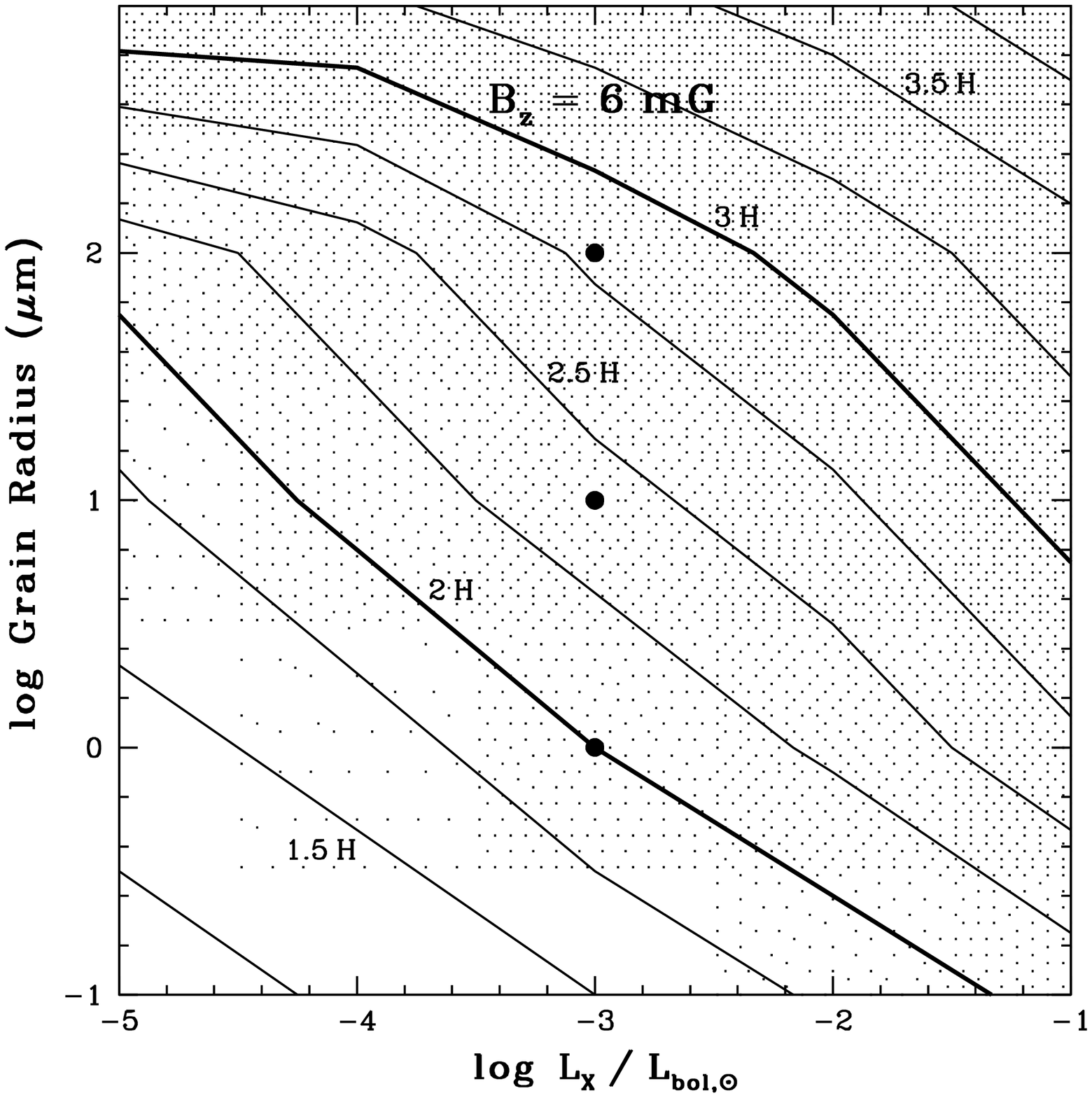}{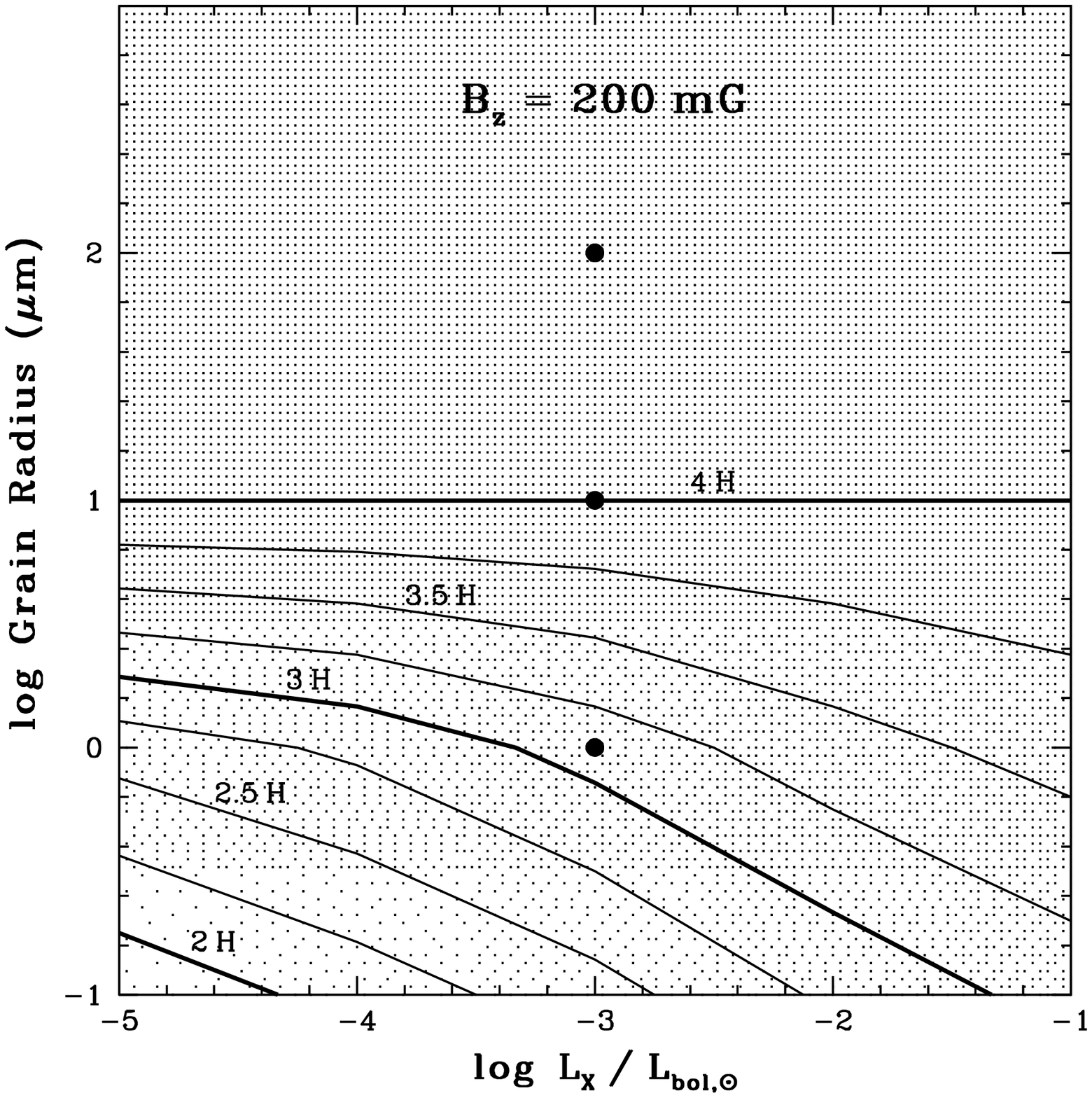}

  \figcaption{\sf Thickness of the active layer as a function of
    stellar X-ray luminosity and grain size, at 5~AU in the
    minimum-mass Solar nebula.  The vertical component of the magnetic
    field is 6~mG (left) and 200~mG (right).  In both panels, the unit
    of luminosity is the solar bolometric luminosity and the shading
    shows the same information as the contours, which are spaced by
    $0.25H$.  The limiting dust-free case appears at the top, taking
    the place of the results for grain radius 1~mm.  Circles mark the
    parameters of our three resistive MHD
    calculations.  \label{fig:activeht}}
\end{figure}

The allowed range of heights for the base of the active layer under
more general magnetic field conditions can be found from
eq.~\ref{eq:elsasser} using the known resistivity profile.  A maximum
height for the base and thus a minimum total thickness for the active
layer comes from considering only the uniform part of the magnetic
field $B_1=6$~mG in the MHD calculations.  A minimum height for the
base and an upper bound on the active layer thickness come from using
the field strength $B_2=200$~mG found in the ideal-MHD tests.  The
resulting limits on the active layer thickness are shown in
figure~\ref{fig:activeht} as functions of the grain size and X-ray
luminosity.  The X-ray luminosity range plotted corresponds to the
spread found in Chandra X-ray Observatory studies of the Orion nebula
that detected most of the lightly-obscured young Solar-mass stars
\citep{gf00,pk05}.  The active layer thickness varies from $1.25H$,
when the field takes the weaker value, the X-ray luminosity is low and
the grains are small, to the full $4H$ extent of the disk, when the
field is strong and the grains are 10~$\mu$m or larger.  The range in
active layer thickness is thus a factor 3.2.

\begin{figure}[tb!]
  \epsscale{0.5}
  \plotone{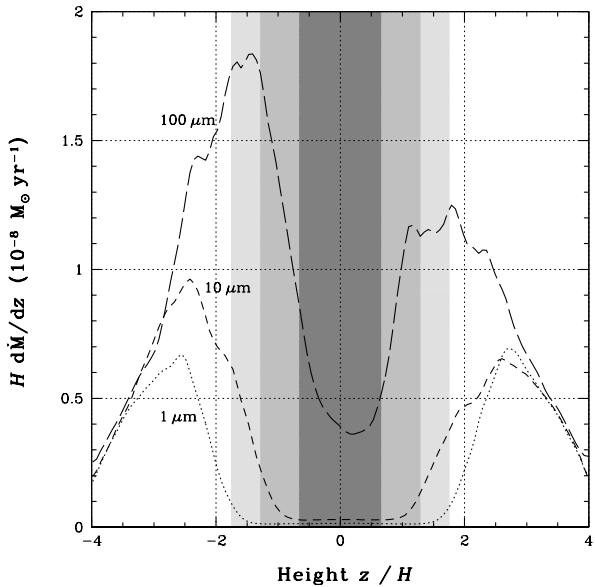}

  \figcaption{\sf Mass accretion rate versus height in three
    shearing-box MHD calculations with settling grains of 1 (dotted),
    10 (dashed) and 100~$\mu$m (long-dashed curve).  The accretion
    rate per unit height is proportional to the stress, which has been
    averaged over the time period from~10 to~60 orbits.  The total
    accretion rates are $2.0\times 10^{-8}$, $3.0\times 10^{-8}$ and
    $7.1\times 10^{-8}$ $M_\odot$~yr$^{-1}$, respectively.  Gray
    strips mark the dead zones where the Elsasser number $Oh$ is less
    than unity.  The darkest gray corresponds to the run with the
    100~$\mu$m particles and the lightest to the 1~$\mu$m particles.
    \label{fig:hdmdotdz}}
\end{figure}

Determining the range of possible mass flow rates involves estimating
the mean stress in the active layers.  The turbulence regenerates the
fields, so that the saturated field strength is determined jointly by
the imposed magnetic flux and, through the resistivity of the gas, by
the X-ray illumination and the distribution of the grains.  The mass
flow and the extent of the active layers in the three MHD calculations
are shown in figure~\ref{fig:hdmdotdz}.  The quantity plotted ${d{\dot
    M}/dz} = 2\pi w_{xy}/\Omega$ is proportional to the total
accretion stress resulting from magnetic and hydrodynamic forces.  The
overall mass flow rate in the run with 100-$\mu$m grains is 3.55~times
that in the run with 1-$\mu$m grains.  This total is made up of a
factor 1.49 due to the range in active layer thicknesses, and a
remaining factor 2.38 due to the stronger magnetic fields found in the
thicker active layers.  If the stresses have the same proportionality
to the active layer thickness over the whole of
figure~\ref{fig:activeht}, then variations in the X-ray luminosity and
grain size can change the mass flow rate by about a factor~16.
Magneto-rotational turbulence can explain the two-decade range in the
observed accretion rates if the remaining factor of six comes from
other effects.  Possibilities include (1) some disks having an
abundance of very small grains, yielding a very thin active layer
\citep{sm00}; (2) a range of magnetic fluxes among T~Tauri systems
[the stress increases with the vertical magnetic flux
\citep{hg95,si04,pc07}]; and (3) changes in the mass flow rate with
radius and time \citep{g96,al01}.

It is of interest to compare the mass column of the active layer with
the X-ray attenuation column, 8~g~cm$^{-2}$.  The active column is
less than the attenuation column when the grains are small and
recombination is fast, and greater when the grains are large.  The
mean active column in the run with 1-$\mu$m grains is 5.18 and 4.82
g~cm$^{-2}$ on the disk top and bottom, respectively.  The numbers are
13.3 and 12.7 g~cm$^{-2}$ in the run with 10-$\mu$m grains, and 32.3
and 33.6 g~cm$^{-2}$ in the run with 100-$\mu$m grains.  The average
values of 5.0, 13.0 and 32.9 g~cm$^{-2}$ span a factor 6.6.

\section{GRAIN MOVEMENTS\label{sec:movements}}

The dust particles in our three MHD calculations move by settling and
by advection with the gas.  The gas is near magneto-hydrostatic
equilibrium throughout.  Its velocity field consists of
magneto-rotational turbulence superimposed on a gradual expansion due
to outflow from the disk surfaces.  Wave motions replace the
turbulence inside the dead zone.  The three following sections outline
the effects on the dust of the settling, the outflow and the turbulent
stirring, while the fourth section deals with the movements of the
particles in the dead zone.

\subsection{Settling\label{sec:settling}}

Dust settles fastest in the upper atmosphere, where the gravity is
strong and the low density means weak gas drag.  In the absence of
turbulence, settling is quick.  According to the rough estimate of
eq.~\ref{eq:tsett}, particles 1, 10 and 100~$\mu$m in size in
100~orbits reach heights of 3.2, 2.4 and $1.1H$, while the more
accurate numerical integrations shown in figure~\ref{fig:zsett} yield
2.5, 1.7 and $0.7H$.  Coagulation can decrease the cross-sectional
area per unit mass, making settling faster still, but we have assumed
here that the particles do not stick to one another.

\begin{figure}[tb!]
  \epsscale{0.5}
  \plotone{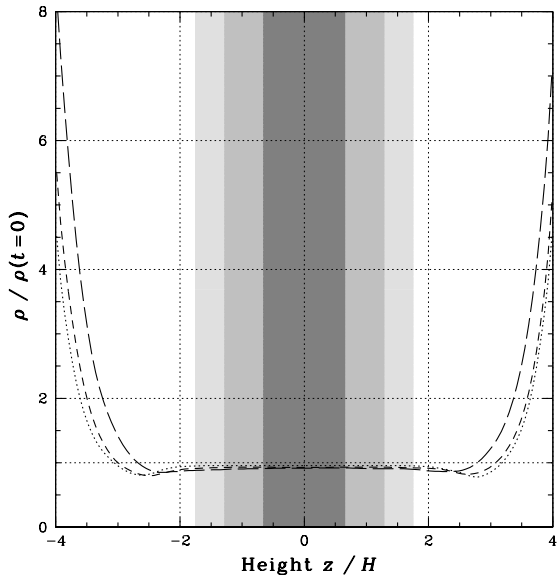}

  \figcaption{\sf Gas density enhancement versus height in the three
    MHD calculations with grains of 1 (dotted), 10 (dashed) and
    100~$\mu$m (long-dashed curve).  The density is scaled by the
    initial hydrostatic equilibrium profile, and averaged horizontally
    and over time from 10 to 60~orbits.  Gray shadings mark the dead
    zones in the three calculations, as in figure~\ref{fig:hdmdotdz}.
    \label{fig:densitiez}}
\end{figure}

The settling speed is inversely proportional to the gas density
(eqs.~\ref{eq:tdrag} and~\ref{eq:vterminal}).  The magnetic fields
cause substantial departures from the hydrostatic equilibrium Gaussian
density profile outside about $2H$, where the magnetic pressure
approaches or exceeds the gas pressure.  Magnetic support allows
higher densities in the upper atmosphere, with the extra material
coming from the deeper layers where the density is correspondingly
reduced (figure~\ref{fig:densitiez}).  The mean gas density at $4H$ is
up to eight times greater than the hydrostatic value.

\begin{figure}[tb!]
  \epsscale{0.5}
  \plotone{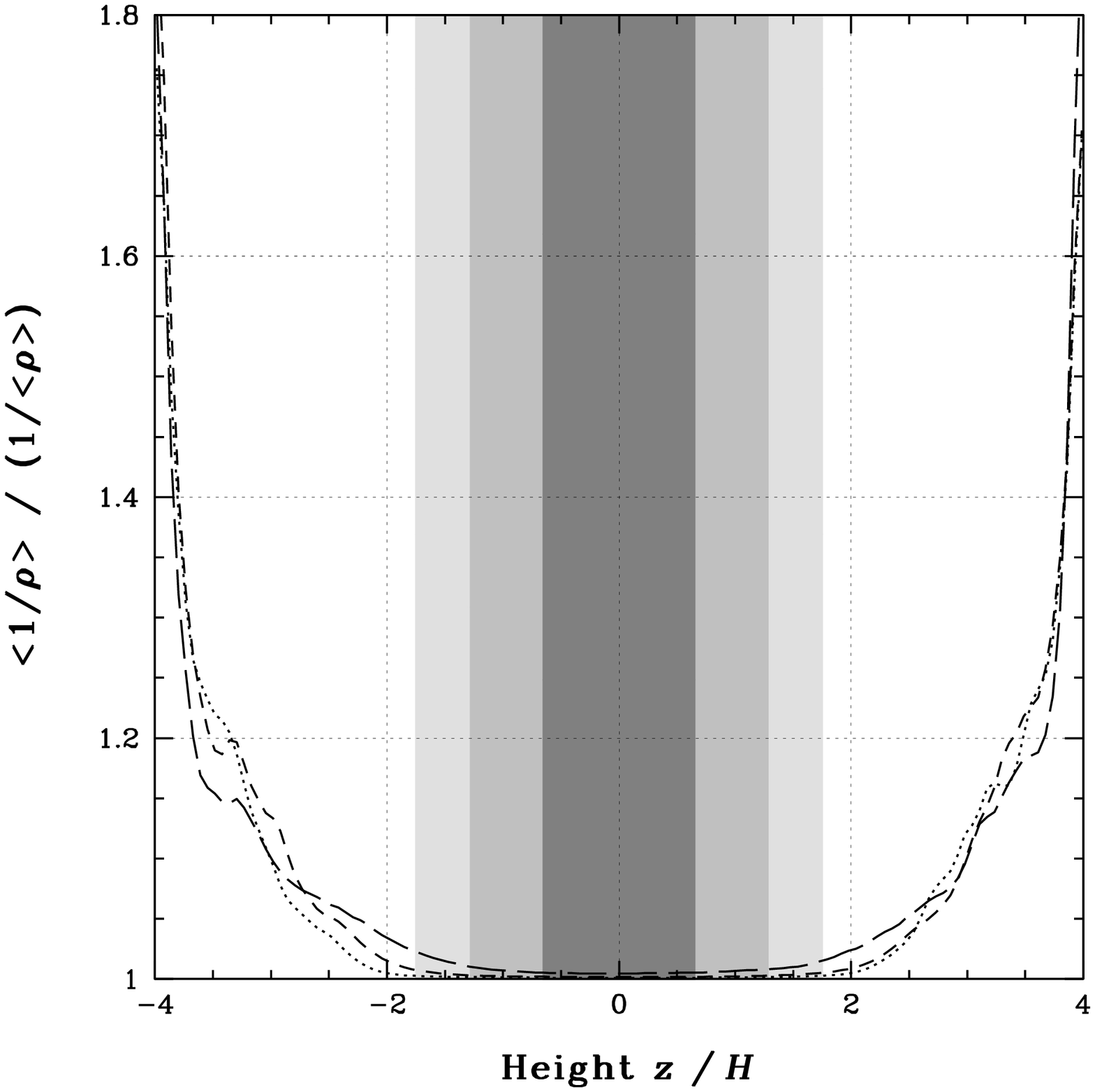}

  \figcaption{\sf Relative increase in grain settling speed due to
    horizontal density fluctuations, versus height in the three MHD
    calculations with grains of 1 (dotted), 10 (dashed) and 100~$\mu$m
    (long-dashed curve).  The averaging indicated by the angle
    brackets is carried out across horizontal surfaces and over time
    from 10 to 60~orbits.  Gray shadings mark the dead zones in the
    three calculations, as in figure~\ref{fig:hdmdotdz}.
    \label{fig:vsettratio}}
\end{figure}

Horizontal density fluctuations in the turbulence however yield an
average settling speed $\sim \left<\rho^{-1}\right>$ greater than the
speed predicted from the mean density $\sim \left<\rho\right>^{-1}$
(figure~\ref{fig:vsettratio}).  The fluctuations make settling faster
than the mean density profile by about 10\% at $3H$ and 80\% at $4H$.
Density fluctuations in the dead zone are small and have little effect
on the settling.

The ionization remains near its time-varying equilibrium value in the
MHD calculations as the grains settle out.  Chemical equilibration
occurs faster than settling except with 100~$\mu$m grains above
3.5~scale heights (figures~\ref{fig:tstop} and~\ref{fig:teqm}).  At
the low densities found above this height, reactions on the surfaces
of the large grains in any case are negligible and the ionization
state is determined by gas-phase processes.

\subsection{Outflow\label{sec:outflow}}

Gas intermittently escapes from the top and bottom of the domain as
the turbulence carries material across the boundaries.  The greatest
speeds are a few times the sound speed, and thus the lost gas remains
gravitationally bound to the star and may return to the disk.  Since
the calculations do not reach high enough nor include the radial
gradients that may be important for any acceleration through escape
speed, we here describe the outflows only to the extent needed to
subtract their effects.

\begin{figure}[tb!]
  \epsscale{0.5}
  \plotone{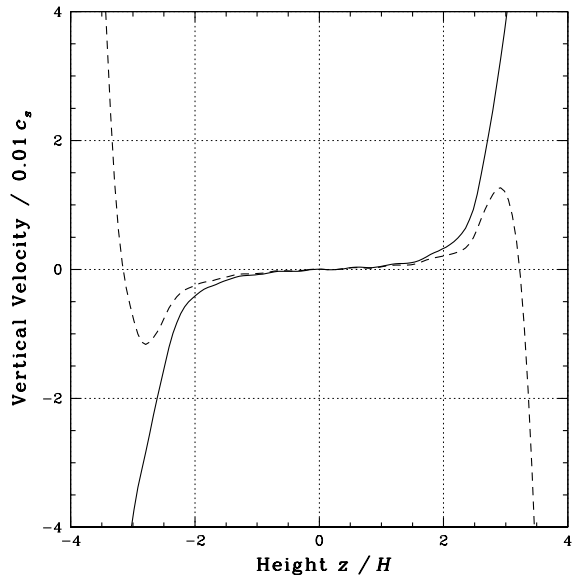}

  \figcaption{\sf Mean vertical velocities of gas (solid line) and
    grains (dashed line) versus height in the MHD calculation with
    10-$\mu$m grains.  The gas motion results from expansion as
    material flows off the disk surfaces.  The grain velocity plotted
    is the sum of the gas velocity and the settling speed, and does
    not include the effects of the turbulence.  Grains below $3.2H$
    are entrained in the outflow, while grains higher in the
    atmosphere settle out.
    \label{fig:settlez}}
\end{figure}

The outflows result in the surface density declining between 10 and
60~orbits by 6.9, 8.3 and 11\% in the calculations with 1-, 10- and
100-$\mu$m grains.  The average mass flow rate in the 1-$\mu$m case is
$2.1\times 10^{-9}$ $M_\odot$ yr$^{-1}$ AU$^{-2}$.  The remaining gas
stays near hydrostatic equilibrium by expanding slightly, and the
expansion carries gas and in some cases dust from the interior into
the disk atmosphere.  For continuity of mass, the expansion speed is
approximately proportional to the inverse density.  The speeds of gas
upwelling and grain settling thus have similar functional increases
with height.  At $2H$, the gas flows away from the midplane at about
0.3\% of the sound speed, while grains of 1, 10 and 100~$\mu$m settle
at 0.012, 0.12 and 1.2\% of the sound speed in the initial Gaussian
density profile according to eq.~\ref{eq:vterminal} and
figure~\ref{fig:tstop}.  The settling is faster than the outflow at
all heights in the run with 100-$\mu$m grains, so that the dust
distribution results mostly from the competition between settling and
turbulent stirring.  By contrast the outflow is faster than the
settling everywhere in the run with 1-$\mu$m grains, but at $4H$ the
grains settle at 30\% of the gas upwelling speed.  As a result, the
grains are lost at a slower rate than the gas and their mass fraction
increases over time in the upper atmosphere.  The run with 10-$\mu$m
grains is an intermediate case, and the outflow carries the particles
to about $3.2H$ (figure~\ref{fig:settlez}).  Above this height the
settling is faster than the outflow.

\subsection{Turbulent Stirring\label{sec:stirring}}

Turbulence causes mixing, making the dust-to-gas ratio more uniform.
The timescale for mixing through one density scale height is
\begin{equation}
  t_M = {H^2 \over t_C \left<(v_z-\overline{v_z})^2\right>},
\end{equation}
where the denominator is the turbulent diffusion coefficient and is
proportional to the correlation time of the turbulence $t_C$ and to
the mean squared turbulent vertical velocity \citep{fp06}.  The
turbulent speed is obtained by subtracting the horizontally-averaged
outflow speed $\overline{v_z}$ from the total.  The mixing timescale
is tens of orbits in much of the active layer.  In particular, at the
dead zone boundary the average mixing time is 24, 53 and 46 orbits in
the calculations with 1, 10 and 100~$\mu$m grains, respectively.  By
comparison, grains initially in the upper atmosphere settle to the
dead zone boundary in 900, 270 and 112 orbits, respectively.  Also,
the outflowing gas crosses one scale height near the dead zone
boundary in approximately 43, 162 and 288 orbits.  The turbulent
mixing is thus the fastest of the three processes at the edge of the
dead zone.

\begin{figure}[tb!]
  \epsscale{0.5}
  \plotone{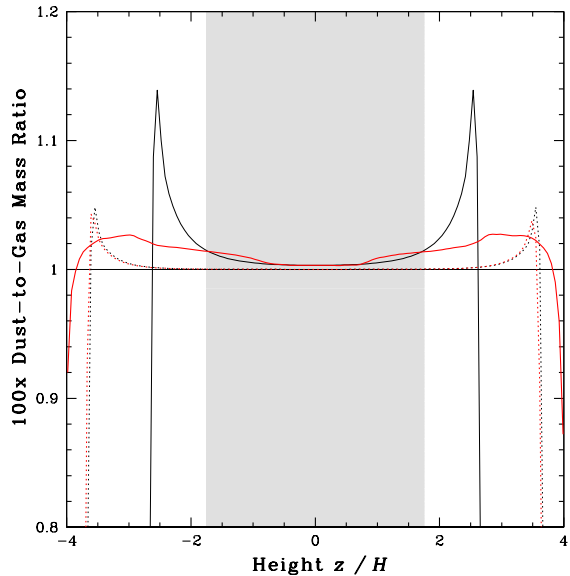}

  \figcaption{\sf Mass fraction of 1~$\mu$m grains versus height.  The
    solutions for a laminar disk at 1 and 60~orbits are shown by thin
    black dotted and solid lines, while the corresponding thick red
    lines are from the shearing-box MHD calculation.  Gray shading
    indicates the dead zone and a thin horizontal line marks the
    initial well-mixed dust abundance.
    \label{fig:xgvsz1um}}
\end{figure}

The results of mixing the 1-$\mu$m grains appear in
figure~\ref{fig:xgvsz1um}, showing the dust mass fraction versus
height.  The dust settles in one orbit to $3.7H$.  By 60~orbits,
settling without turbulence produces a spike in the dust abundance
near $2.7H$.  In the MHD calculation where outflow and turbulence
occur alongside the settling, the spike is smoothed out across the
active layer as follows.  The settling enhances the dust abundance at
2 to $3H$.  The outflow carries some of the dust-enriched material
upward, and the turbulence spreads the dust-enriched material both up
into the atmosphere and down into the dead zone.  The downward mixing
overshoots the dead zone boundary by about one scale height.  Settling
and turbulence together thus speed up the concentration of the solids
in the deeper layers compared with the laminar case.  This contradicts
the usual wisdom that turbulence counteracts settling.

\begin{figure}[tb!]
  \epsscale{0.5}
  \plotone{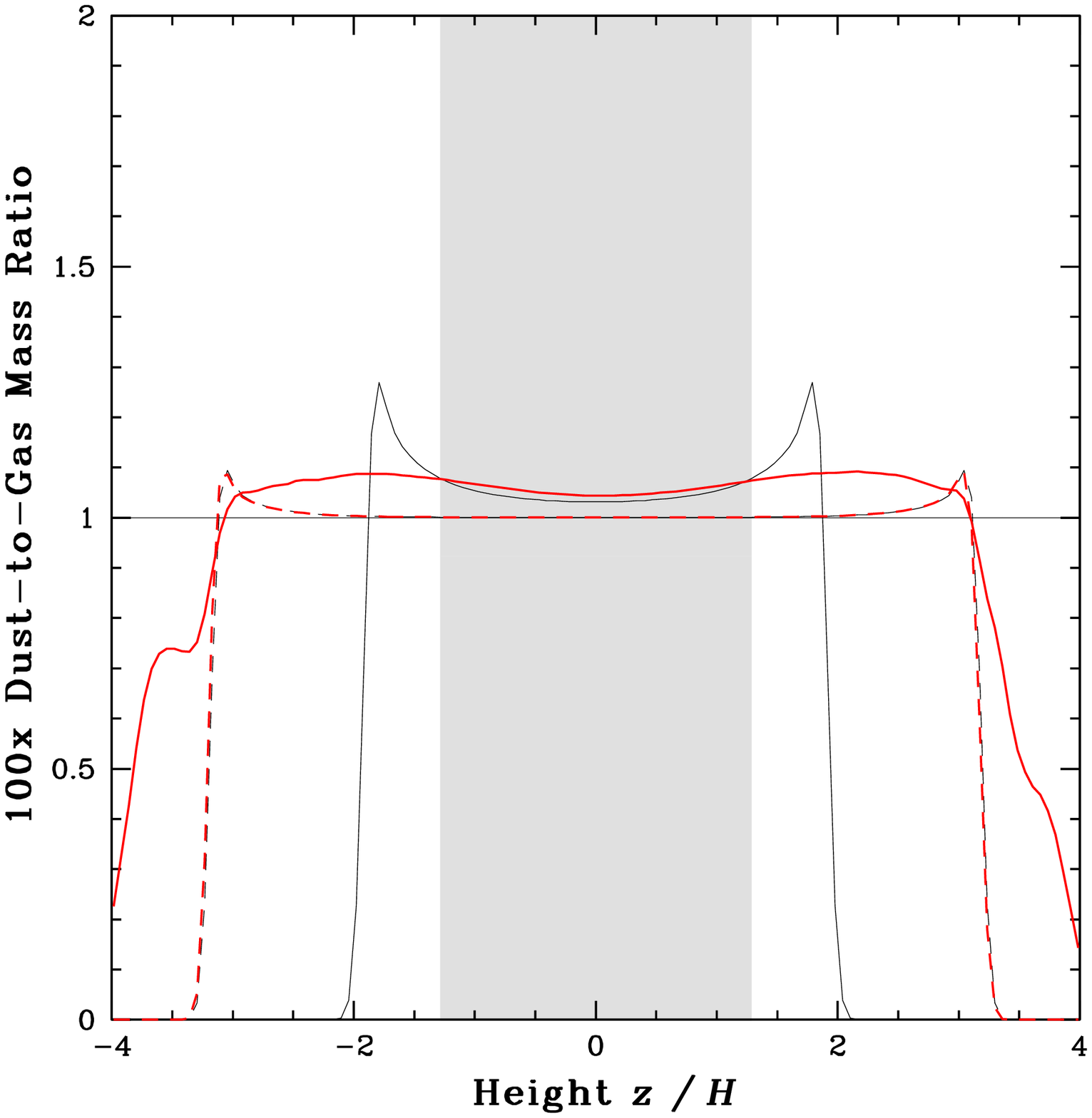}

  \figcaption{\sf Mass fraction of 10~$\mu$m grains versus height.
    The solutions for a laminar disk at 1 and 60~orbits are shown by
    thin black dashed and solid lines, while the corresponding thick
    red lines are from the shearing-box MHD calculation.  Other
    symbols are as in figure~\ref{fig:xgvsz1um}.
    \label{fig:xgvsz10um}}
\end{figure}

Similar processes operate on the 10-$\mu$m grains
(figure~\ref{fig:xgvsz10um}).  Few of the particles remain by
60~orbits above $3.5H$, where the settling time is less than an orbit
and settling overwhelms both the turbulent mixing and the entrainment
in the outflow.  The turbulence mixes the dust almost uniformly
through the dead zone, as the movements of the gas overshoot the dead
zone boundary all the way to the midplane.

\begin{figure}[tb!]
  \epsscale{0.5}
  \plotone{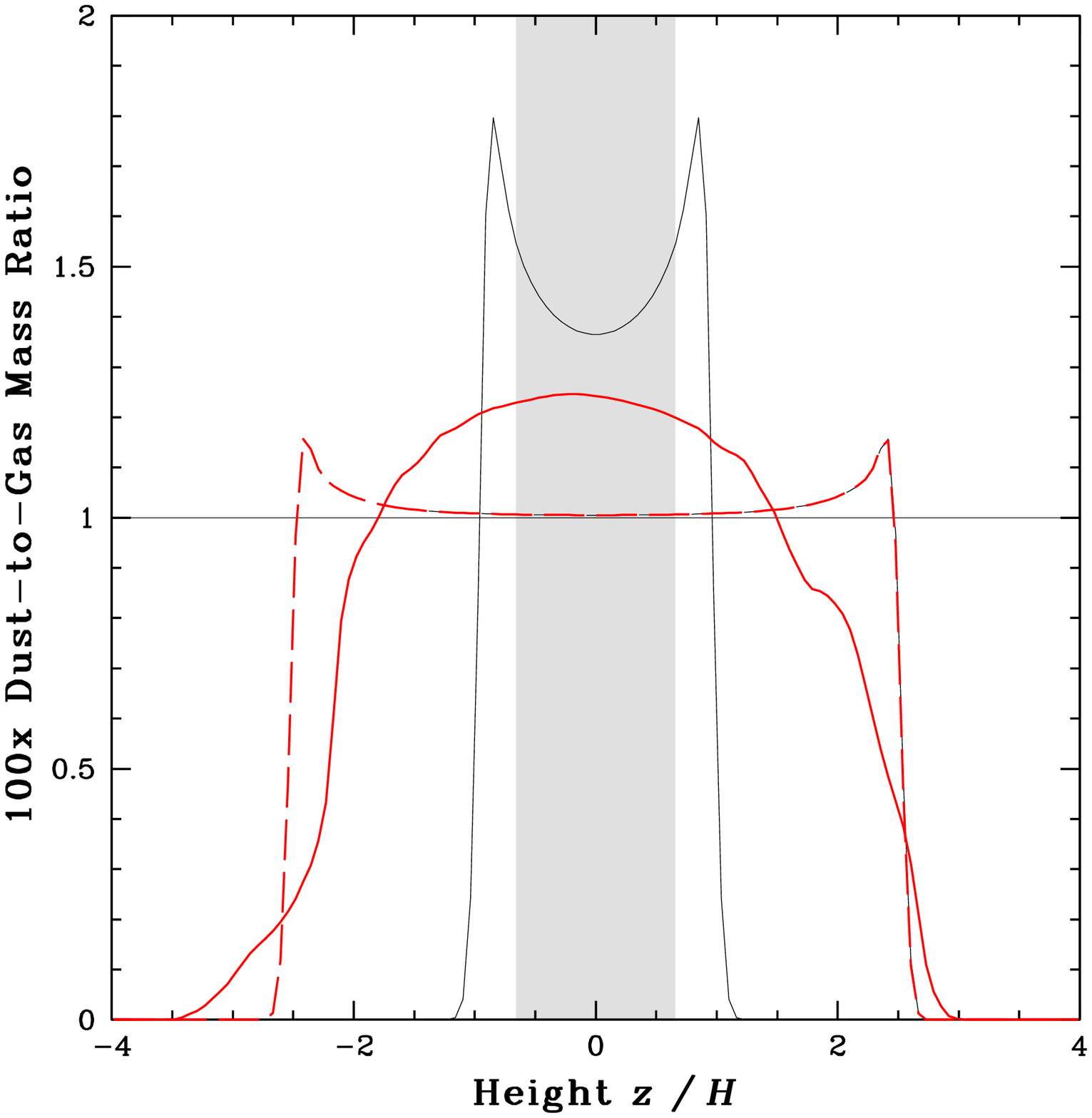}

  \figcaption{\sf Mass fraction of 100~$\mu$m grains versus height.
    The solutions for a laminar disk at 1 and 60~orbits are shown by
    thin black long-dashed and solid lines, while the corresponding
    thick red lines are from the shearing-box MHD calculation.  Other
    symbols are as in figure~\ref{fig:xgvsz1um}.
    \label{fig:xgvsz100um}}
\end{figure}

The 100-$\mu$m grains (figure~\ref{fig:xgvsz100um}) have mostly
settled below $2.5H$ by 1~orbit and remain there at 60~orbits.  Nearer
the midplane, the stirring and settling timescales are comparable.
The dead zone extends just $0.65H$ either side of the midplane, and
overshooting turbulent motions reach all parts of the interior.  The
settling is fast enough that the dust abundance has become greatest
close to the midplane.  The turbulent mixing thus on balance moves
grains from the dead zone to the active layers, preventing further
concentration.

\begin{figure}[tb!]
  \epsscale{0.5}
  \plotone{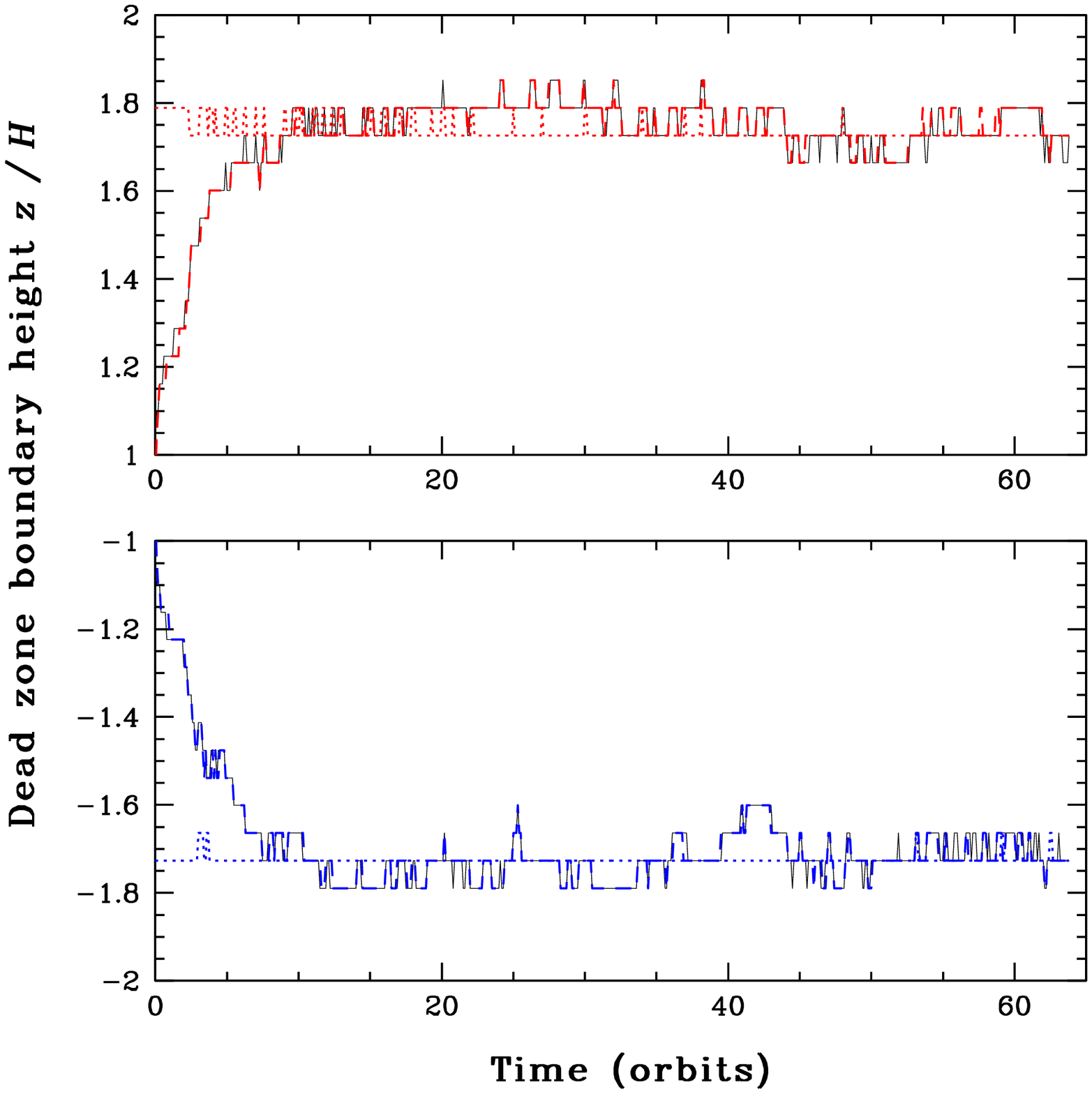}

  \figcaption{\sf Height of the upper and lower boundaries of the dead
    zone in the MHD calculation with 1-$\mu$m grains.  Thin solid
    lines show where the Elsasser number $Oh=v_{Az}^2/\eta\Omega$ is
    unity.  Thick dotted lines show unity for the Elsasser number
    calculated using the time-averaged squared vertical Alfv\'en
    speed.  All of the time variation then comes from the resistivity.
    Thick dashed lines show unity for the Elsasser number calculated
    using the time-averaged resistivity, so all of the time variation
    comes from the Alfv\'en speed.  The dashed lines overlap the solid
    lines, indicating that the changes in Alfv\'en speed account for
    most of the variation.  The expansion of the dead zone over the
    first few orbits results from the dissipation of the initial
    radially-varying magnetic field.
    \label{fig:deadzonebdy}}
\end{figure}

The movements of the dust have little effect on the dead zone depth
over the time interval spanned by the calculations.  In
figure~\ref{fig:deadzonebdy} we plot three different measures of the
dead zone boundary height versus time, in the case with 1-$\mu$m
grains.  The Elsasser number from eq.~\ref{eq:elsasser} is computed
(1) holding the resistivity profile fixed at its time-averaged level,
(2) holding the squared Alfv\'en speed profile fixed, and (3) allowing
both the resistivity and the Alfv\'en speed to vary with time.  The
first and third cases generally coincide, showing that it is the
magnetic field, rather than the rate of recombination on grains, that
controls the changes in the dead zone size over timescales comparable
to the orbital period.

To check whether neglecting particle growth is valid, we compare the
expected grain-grain collision speeds with the threshold of about
1000~cm~s$^{-1}$ for the disruption of silicate aggregates established
by laboratory measurements \citep{bw08}.  Adopting the expressions for
the relative velocities of particles in isotropic turbulence derived
by \cite{oc07}, we find that aggregates of 10~$\mu$m or larger can be
disrupted by collisions with 1-$\mu$m particles in the three MHD
calculations above about $3H$.  Compact spheres as large as
100~$\mu$m settle fast enough that they rarely reach such heights,
judging from figure~\ref{fig:xgvsz100um}.  Meanwhile near the
midplane, collisions are too slow for disruption, and growth to larger
sizes is likely.

\subsection{Grains in the Dead Zone\label{sec:deadzone}}

\begin{figure}[tb!]
  \epsscale{0.5}
  \plotone{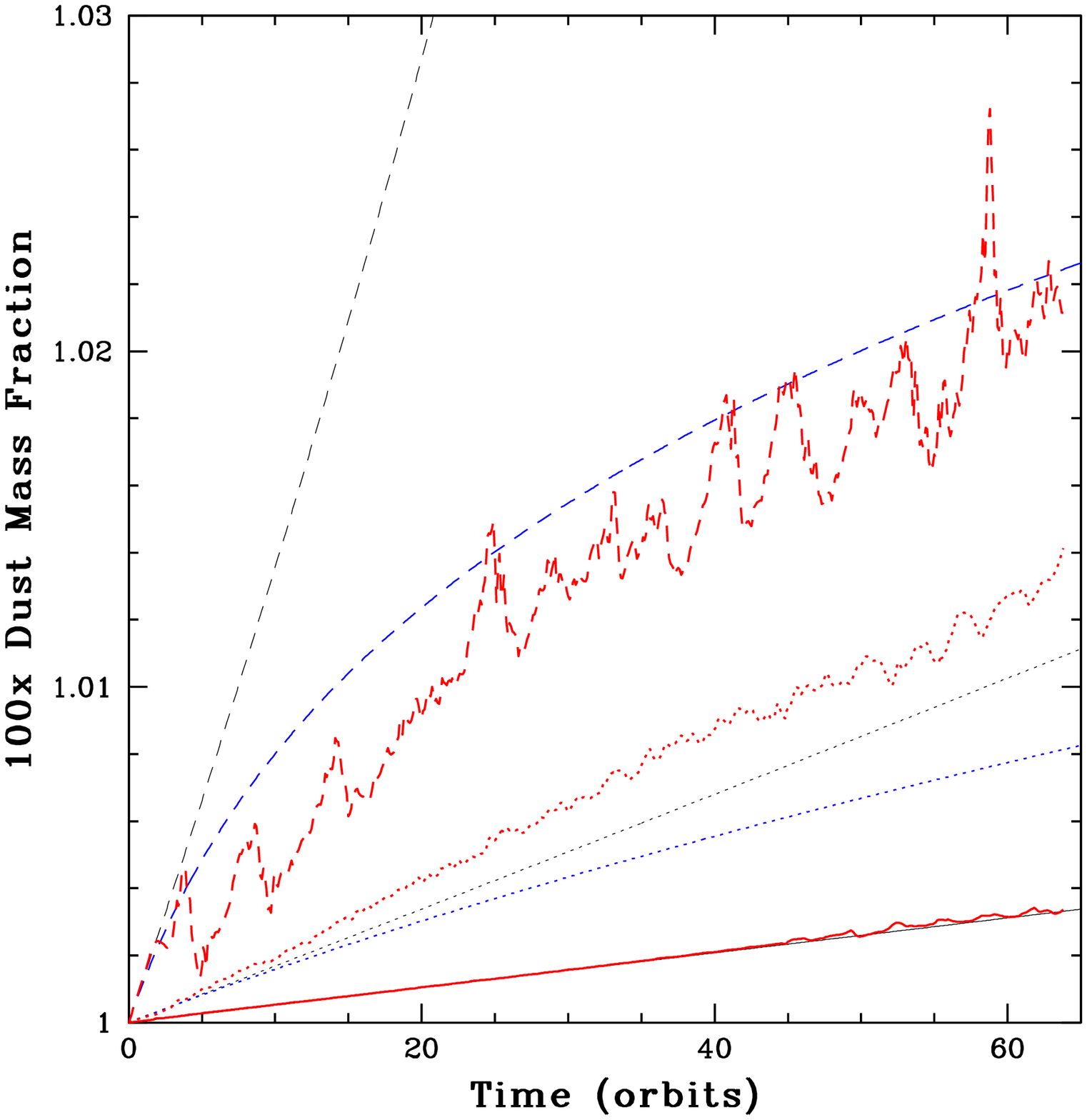}

  \figcaption{\sf Mass fraction of 1~$\mu$m grains versus time, at the
    midplane (solid) and just inside and outside the dead zone
    boundary at 1.5 (dotted) and 2.5 (dashed) scale heights.  The thin
    black lines show the solution for a hydrostatic disk, the medium
    blue lines the solution for a steady laminar outflow, and the
    thick red lines are from the shearing-box MHD calculation.
    \label{fig:deadsett}}
\end{figure}

In this section we examine the movements of the dust particles in and
near the dead zone.  We focus first on the calculation with the
1-$\mu$m grains, where the dead zone is deep enough that turbulent
mixing fails to reach the midplane.  To isolate the contributions of
settling and outflow we compare the MHD results against (1) the
hydrostatic settling case and (2) a numerical solution of the dust
continuity equation in a time-constant flow $v_z = 4\times 10^{-4} c_s
\exp\left(z^2/2H^2\right)$ that approximates the outflow found in the
MHD calculation.  The expansion of the gas carries less-settled and
therefore less-dust-enhanced interior material into the surface
layers, leading to slower increases in the dust abundance over time
than in the hydrostatic case.  The evolution of the dust abundance at
several different heights is shown in figure~\ref{fig:deadsett}.
Outside the dead zone boundary, at $2.5H$, the abundance in the MHD
calculation increases and decreases intermittently due to the
turbulence, while overall tracking slightly below the outflow
solution.  The deficit results from the loss of dust to the deeper
layers through turbulent mixing (figure~\ref{fig:xgvsz1um}).  Just
inside the dead zone boundary, at $1.5H$, the abundance is higher
than the outflow solution owing to the dust mixed down from above.
Meanwhile at the midplane, the dust abundance follows the laminar
solution with small oscillations.  Deep in the dead zone, the settling
proceeds largely as if there were no turbulence in the active layers.

\begin{figure}[tb!]
  \epsscale{0.5}
  \plotone{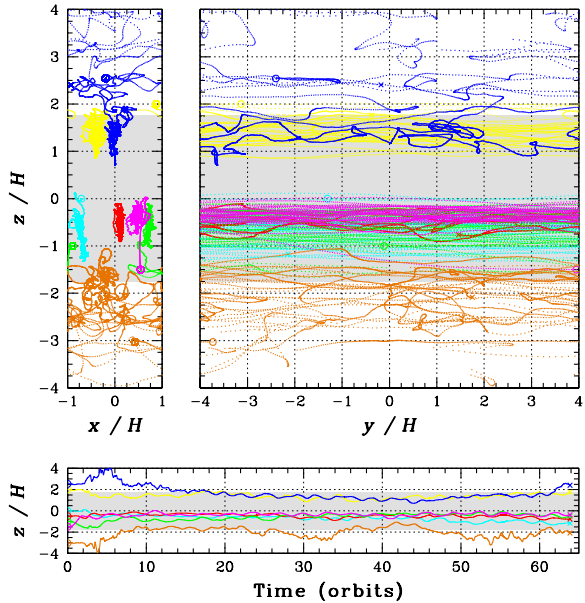}

  \figcaption{\sf Trajectories of seven 1~$\mu$m particles in the MHD
    calculation.  The particle paths are viewed along the orbit (top
    left) and from the star (top right), while the particle heights
    are plotted against time in the bottom panel.  Circles mark the
    initial positions and crosses mark the positions at 64 orbits when
    the calculation ends.  Gray shading indicates the dead zone.
    \label{fig:trajctry}}
\end{figure}

To see how turbulence overshooting the dead zone boundary affects the
motions of individual grains, we have also computed the paths followed
by a large number of particles in the time-dependent gas density and
velocity field of the MHD calculation using the Runge-Kutta scheme
described in section~\ref{sec:laminar}.  Selected particles are shown
in figure~\ref{fig:trajctry}.  The grains found in the active layers
follow looping paths in the turbulence, as on parts of the blue and
orange curves.  The radial and vertical speeds are comparable, and the
vertical motions plotted in the bottom panel are irregular.  Grains
found in the dead zone in contrast show mostly slower periodic
vertical oscillations and have little radial motion, as on the red,
green, cyan and magenta curves.  Particles cross the dead zone
boundary moving both toward and away from the midplane, while
particles near the midplane remain close to their initial positions.
The non-zero hydrodynamic accretion stress in the dead zone results
from propagating waves, which have little impact on the particle
settling.  On the other hand, the particles near the midplane
oscillate with amplitudes a few tenths of a scale height and phases
that drift over time relative to one another, suggesting that the
settling will not yield a thin flat midplane dust layer of the kind
typically assumed for the formation of planetesimals by gravitational
instability \citep{gw73}.  The thickness of the settled layer in an
otherwise stationary nebula is just 1\% of the gas scale height, if
limited by the onset of the Kelvin-Helmholtz instability between the
dust-rich Keplerian component and the overlying gas that orbits at
sub-Keplerian speeds owing to its partial support against the stellar
gravity by the radial gas pressure gradient \citep{cy09}.

\section{TIME VARIABILITY\label{sec:variable}}

Time variations in the strength and shape of the silicate band at
wavelength 10~$\mu$m have now been seen by several groups using
instruments on the ground and in space.  About a quarter of the two
dozen low- and intermediate-mass young stars monitored by \cite{wh00}
and \cite{ww04} from 1992 to 2002 showed significant variations in the
silicate bands but not in the continuum.  The observations were
separated by months to years and the variations showed no clear
period.  The generally poor correlations with changes in optical
brightness and extinction appear to rule out an origin in the
reprocessing of the starlight by a static envelope, but may be
consistent with the features arising from dust in the disk atmosphere
\citep{wb00}.  Judging from static models, the 10-$\mu$m emission
would come mostly from dust near 1~AU \citep{dc06}.  A much larger
census of the Chameleon~I star-forming region with the Spitzer Space
Telescope detected photometric variations at 3.6 to 8~$\mu$m in about
half of all disk-bearing sources observed at intervals of months to
years \citep{la08}.  The silicate band varied on comparable timescales
in two of 11 accreting T~Tauri stars observed with Spitzer by
\cite{bl09} and in the EX~Lupi system observed by \cite{aj09}.  A
further system, the transitional disk LRLL~31, exhibits variability in
the silicate feature and mid-infrared continuum in observations
separated by as little as a week, which might be due to changing
shadows cast on the disk by material near the star \citep{mf09}.  In
several examples, spatially resolved optical and near-infrared images
taken a few days to months apart show changes in shape and brightness
at distances of 50 to 350~AU \citep{bs96,cs05,ws07,lm09}.  Dynamical
processes operating at these radii appear incapable of producing such
rapid changes.  Photopolarimetry of one of the objects, HH30, shows
periodic modulation on timescales of about a week, explained by a
lighthouse picture in which a beam or shadow from a central source
sweeps across the disk \citep{dw09}.  Altogether, the evidence is
consistent with the widespread infrared variability arising at least
in part from changes in the distribution or illumination of the dust
in the disk atmospheres.  To help evaluate whether magnetic activity
moves the dust around enough to cause the variability, we use the
results of the MHD calculation with 1-$\mu$m particles in this section
to compute in turn the height of the photosphere, the appearance of
the disk patch at mid-infrared wavelengths, and the obscuring column
along lines of sight grazing the disk surface.

\begin{figure}[tb!]
  \epsscale{0.5}
  \plotone{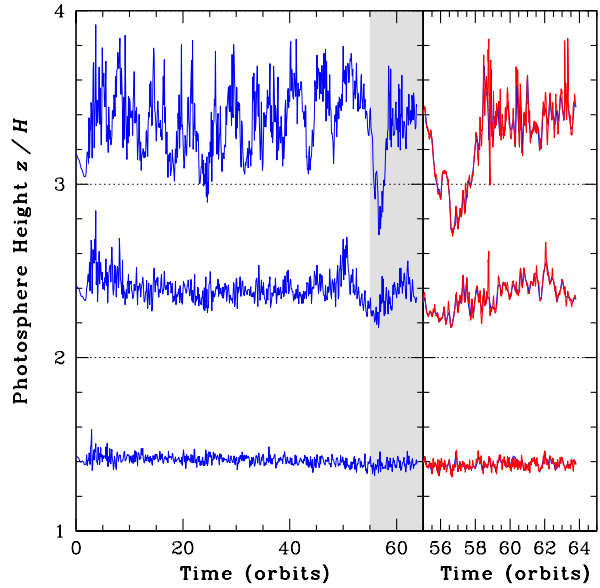}

  \figcaption{\sf Time history of the height where the optical depth
    is unity in the calculation with 1-$\mu$m grains.  The optical
    depth is measured along the line of sight passing vertically down
    through domain center.  Three opacities are considered: 1000, 100
    and 10~cm$^2$ per gram of dust (top to bottom).  The measurements
    of the optical depth are made ten times per orbit.  In the
    right-hand panel the last nine orbits, marked by a gray band in
    the main figure, are shown on an expanded time scale.  The MHD
    calculation was restarted with 100 optical depth measurements per
    orbit and the results are overplotted as a thin red line.  At the
    highest opacity the photosphere height ranges from 2.7 to 3.8$H$,
    considering the period after 10~orbits when the turbulence is well
    established.  With opacity ten times less, the height ranges from
    2.2 to 2.7$H$ and with opacity a hundred times less, the range is
    1.31 to 1.48$H$.
    \label{fig:tau1}}
\end{figure}

The photosphere height is plotted versus time in
figure~\ref{fig:tau1}.  Optical depths are calculated by integrating
the dust column along the ray passing vertically down through the
center of the box.  The results are shown for opacities of 1000, 100
and 10~cm$^2$ per gram of dust.  The highest of these is comparable to
the geometric cross-section of 1500~cm$^2$~g$^{-1}$ for the 1-$\mu$m
grains and to the 982~cm$^2$~g$^{-1}$ opacity at wavelength 10~$\mu$m
obtained for grains in dense molecular cloud cores by \cite{po93}.
This highest opacity places the photosphere in the most
magnetically-active layers, where the fluctuations in the dust density
are greatest.  The location of the photosphere varies over time by
more than a scale height.  The lower-opacity curves correspond to dust
depletions of one and two orders of magnitude.  At the lowest opacity,
the active layers are optically thin and the photosphere lies inside
the dead zone.  Optically-thin active layers at 1~AU have been
suggested as a consequence of the low dust abundance needed for
magnetic stresses there to yield the measured stellar accretion flows
\citep{bg09}.

\begin{figure}
  \epsscale{1}
  \plotone{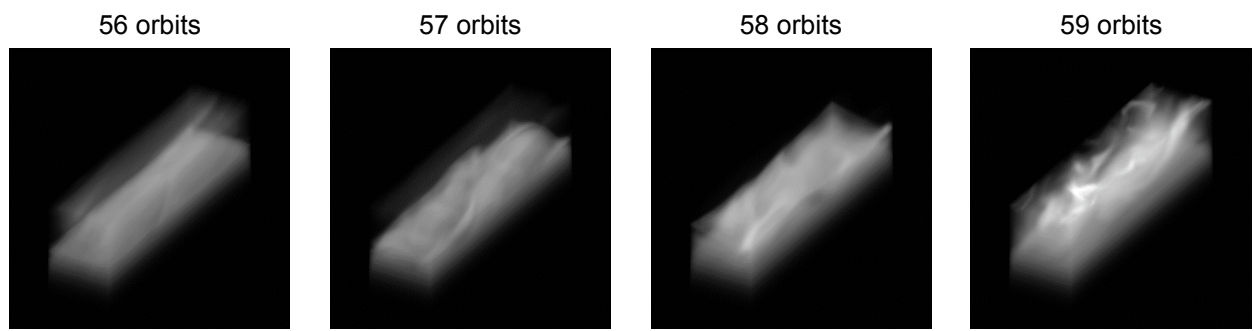}

  \figcaption{\sf Time sequence of images at wavelength 10~$\mu$m
    showing the thermal emission from the dust in the MHD calculation
    with 1-$\mu$m grains.  The magnetic field grows stronger from 56
    to 58~orbits, reducing the gas density so that the dust quickly
    settles out and the optical depth of the disk atmosphere declines.
    Around 58~orbits, buoyancy expels the magnetic field from the
    disk.  The denser gas below expands into the empty layer, sweeping
    dust back into the atmosphere, so that the optical depth and
    surface brightness increase.
    \label{fig:movie}}
\end{figure}

\begin{figure}
  \epsscale{1}
  \plotone{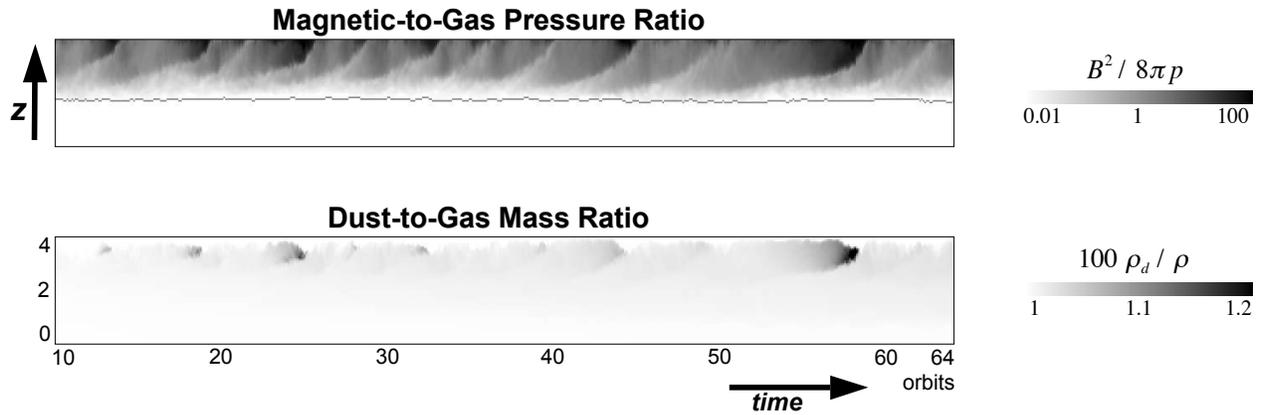}

  \figcaption{\sf Ratios of magnetic to gas pressure (top panel) and
    dust to gas mass density (bottom panel) versus height and time in
    the MHD calculation with 1-$\mu$m grains.  Only the top half of
    the domain is shown, and both quantities are
    horizontally-averaged.  A gray line through the middle of the top
    panel marks the dead zone boundary where $Oh=1$.  Episodes of
    strong magnetic fields force the gas to expand, leading to low gas
    densities that allow grains to quickly become concentrated through
    settling.
    \label{fig:episodic}}
\end{figure}

The time variation in the thermal emission from the dust is
illustrated in figure~\ref{fig:movie}.  We calculate the emission at
wavelength 10~$\mu$m using a ray-tracing radiative transfer code based
on that of \cite{y86} as described by \cite{tb97}, taking an opacity
of 100~cm$^2$ per gram of dust.  Scattering provides a small fraction
of the total opacity \citep[e.g.][]{po93} so we treat only absorption
and thermal emission.  To mimic the warm surface layer produced by
starlight \citep{cg97}, stellar X-ray irradiation \citep{gn04} and
photoelectric heating \citep{kd04}, we assume when solving the
transfer equation that the temperature increases linearly above one
scale height, reaching 600~K at $4H$.  Temperature fluctuations due to
changes in the location where the stellar radiation is absorbed are
neglected.  The time sequence in figure~\ref{fig:movie} shows the
brightness variations over a three-orbit span.  Every few orbits, the
magnetic fields along the orbital direction grow strong through the
shear, reducing the gas density by one to two orders of magnitude so
that the dust quickly settles out (figure~\ref{fig:episodic}).  The
fields eventually become buoyant and rise through the boundary.  Gas
expanding into the space vacated by the fields carries dust particles
back into the warm layer where they become bright in the 10-$\mu$m
band.  The eruptions of dusty gas recur episodically, synchronized
with the variations in the magnetic activity.

\clearpage
\clearpage

Uncorrelated brightness variations in many patches across the disk
will produce only slight changes in the emission of the system as a
whole.  However the atmosphere can contribute to variability also by
casting time-dependent shadows on the more distant parts of the disk,
and by obscuring the central star in systems viewed near edge-on.  The
shadows sweep across the disk on the orbital period of the shadowing
material, while possibly changing in size and strength over shorter
timescales determined by the turbulence.  We estimate the magnitude of
the faster variations by measuring the mass column along a horizontal
ray that is fixed in the rotating frame of the shearing-box, passing
$3.75H$ above the origin.  The extinction to the central star as seen
by an outside observer, on the other hand, changes as the orbital
motion carries material through the line of sight.  We estimate the
timescales involved by measuring the mass column along a ray that is
fixed in the inertial frame and moves across the shearing-box at the
orbital speed, making use of the periodic azimuthal boundary
condition.  The inertial line of sight crosses the azimuthal extent of
the domain in just 0.06~orbits.

\begin{figure}[tb!]
  \epsscale{1}
  \plottwo{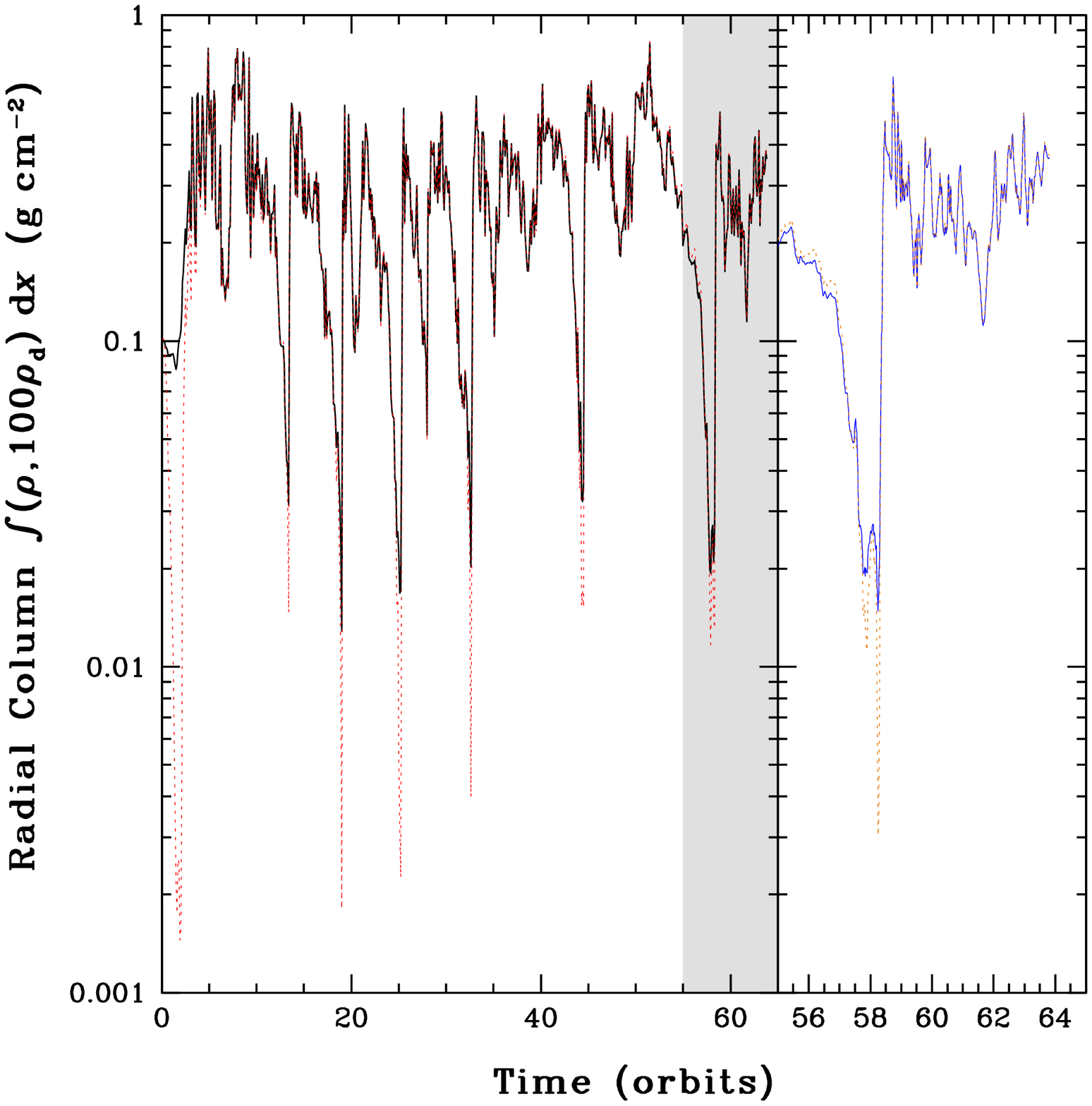}{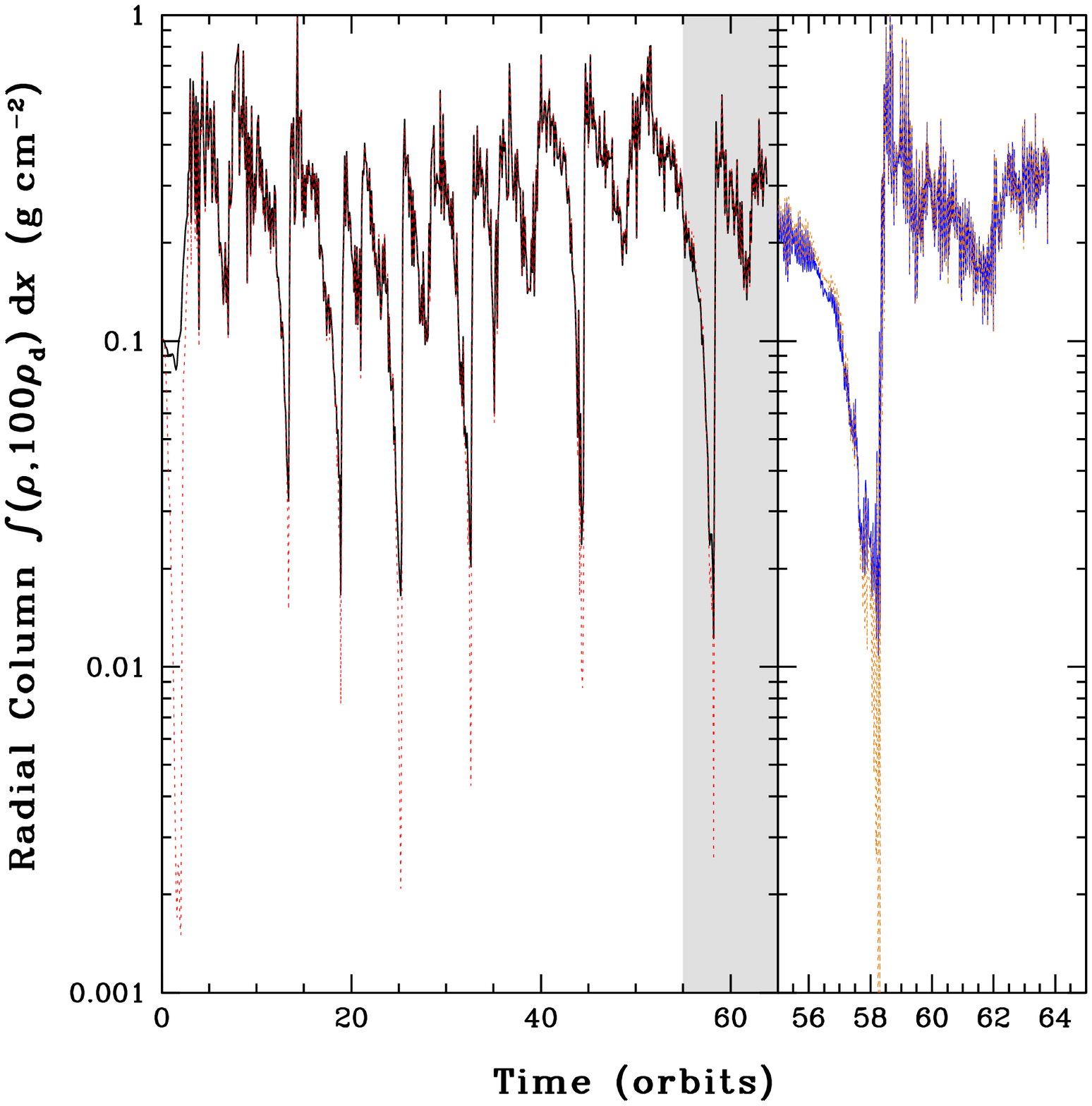}

  \figcaption{\sf Time history of the mass column along a radial ray
    $3.75H$ above the midplane in the calculation with 1-$\mu$m
    grains.  At left, the line of sight is fixed in the rotating frame
    relevant for changes in the shadows cast on the outer disk.  At
    right, the line of sight is fixed in the observer frame, relevant
    for changes in the stellar extinction in systems viewed through
    the disk atmosphere.  In both panels the gas column is drawn in
    solid black and the dust column, scaled up by a factor 100, is
    drawn in dotted red.  The measurements are separated by an
    interval of 0.1~orbit.  A gray shaded band marks the time interval
    where the MHD calculation was repeated with measurements taken
    every 0.01~orbit.  The results are plotted to the right of each
    panel on an expanded time scale in solid blue (gas) and dotted
    orange (dust).  For the rotating line of sight, ten measurements
    per orbit is sufficient to capture most of the time variation.  By
    contrast, as seen at right the orbital motion of the density
    fluctuations through the observer's line of sight causes
    flickering on timescales down to at least 0.01~orbit.
    \label{fig:xcolumn}}
\end{figure}

The columns of gas and dust are plotted against time in
figure~\ref{fig:xcolumn}.  Over intervals of several orbits, the
column decreases by an order of magnitude due to the growth of the
magnetic field, before quickly returning to its old level when the
field escapes.  The dust settles out fastest when the column is low,
reducing the ratio of dust to gas.  Faster, erratic column variations
on timescales $t_C\approx 1/\Omega\approx 0.16$~orbit result from the
density fluctuations in the turbulence, while still faster variations
over about 0.01~orbit, visible only on the inertial line of sight,
result from the orbital motion.  As seen on the rotating line of
sight, the dust column changes by more than a factor two in 0.1~orbit
every 2.3~orbits on average, while similar changes within a shorter
interval of 0.01~orbit occur just once in 8.8~orbits.  On the inertial
line of sight, dust column measurements separated by 0.1~orbit differ
by more than a factor two every 1.4~orbits, and those separated by
0.01~orbit show a similar change every 0.22~orbits.  These results
indicate that the turbulence will yield changes in the depth of the
shadows cast on the disk mostly over timescales longer than a tenth of
an orbit, while changes in the extinction to an observer will be
substantial on timescales down to at least 0.01~orbit.  Note that the
patch of disk treated in the MHD calculation has radial optical depth
near unity at $3.75H$ and thus can cast a strong shadow, if the
opacity is a few hundred cm$^2$ per gram of dust.

In the calculation with 10-$\mu$m grains, the settling is so effective
at times of low column that the upper atmosphere becomes almost
dust-free.  The gas expands owing to the magnetic pressure making the
densities less than the initial hydrostatic profile.  The particles
then settle below $3.5H$ in less than an orbit
(figure~\ref{fig:xgvsz10um}).  The case with 100-$\mu$m grains is
dust-free at $3.75H$ almost always.

\section{DISCUSSION \& CONCLUSIONS\label{sec:conc}}

We have made resistive MHD calculations of a patch of disk in orbit
around a young star, treating simultaneously for the first time three
processes that together regulate the distribution of
magneto-rotational turbulence.  First, the dust drifts through the gas
under the gravity and gas drag forces.  Second, the dust alters the
resistivity of the gas, and third, the resistivity controls the
evolution of the magnetic fields driving the gas flows.  The three
processes combine in quite different ways in the disk atmosphere and
interior.

In the atmosphere, the high flux of ionizing radiation from the star
allows good coupling between gas and magnetic fields even when the
dust abundance is high.  Intermittency in the turbulence and in the
distribution of dust result from fluctuations in the magnetic field
strength.  During intervals of weak fields, the magnetic forces drive
turbulence that carries dust particles high in the atmosphere.  The
turbulent mixing of the grains is faster than either the settling or
the outflows resulting from the open boundaries.  When the fields are
stronger, their large pressure prevents magneto-rotational turbulence
and reduces the gas density, so that the particles settle quickly.

At the same time in the dead zone near the disk midplane,
recombination on grain surfaces reduces the ionization fraction below
the level needed to sustain turbulence.  Grains settle at the laminar
rate, with superimposed vertical oscillations due to the propagation
of waves excited in the turbulent layers.  Particles falling into the
laminar layer do not return to the disk atmosphere and will eventually
concentrate near the midplane, in a layer whose thickness is not less
than the minimum imposed by the vertical oscillations.  A dead zone
enriched in solids may provide a favorable environment for the growth
of larger bodies.  Many T~Tauri systems show a combination of
10-$\mu$m silicate emission and shallow millimeter spectral slope that
is consistent with micron-sized grains in the disk atmosphere and
millimeter-sized or larger particles in the interior, though it should
be noted that the two wavelengths probe different distances from the
star \citep{dc06,pp08}.

The turbulent and dead zones exchange dust and gas when turbulent
motions overshoot the mutual boundary.  The mixing extends about one
scale height into the dead zone, as shown by the penetration of the
dust-enriched material in figures~\ref{fig:xgvsz1um}
to~\ref{fig:xgvsz100um}.  The motions can perhaps be described in
terms of Alfv\'en waves driven by the breakup of channel flows in the
disk atmosphere \citep{si09}.  The net effect is a transfer of dust to
the dead zone, because settling proceeds fastest in the disk
atmosphere.  During the early stages of the concentration of the solid
material, turbulence thus speeds up settling.  The turbulent layer
will eventually become dust-free unless replenished through radial
transport within the disk or rain-out from an outflow or from the
circumstellar envelope.

Well-mixed micron or sub-micron grains with the interstellar mass
fraction produce a dead zone reaching more than two scale heights from
the midplane at 5~AU in the minimum-mass solar nebula.  However,
grains 10~$\mu$m or larger yield a dead zone ending below one scale
height.  Turbulent mixing then penetrates to the midplane, consistent
with the finding by \cite{fp06} that diffusion adequately describes
the effects of the gas motions on particles in a dead zone with a
small vertical extent.  While we were able to run the case with
1-$\mu$m grains only about 60~orbits, the continuing loss of dust into
the dead zone without resupply will ultimately yield a depletion
factor in the surface layers exceeding the value of about ten needed
for mixing to reach the midplane.

The rate at which mass flows through the patch of disk toward the star
is the product of the turbulent layer depth with the mean accretion
stress.  Both these quantities increase with the grain size, the
ionizing flux from the star, and the magnetic flux threading the disk.
Depletion of the dust in the surface layers will have a similar effect
to an increase in the grain size.  The two-decade spread in accretion
rate among T~Tauri stars cannot easily be produced by grain growth and
the observed spread in stellar X-ray output, but can be understood if
in addition the objects have a modest range of disk magnetic field
strengths.  The mass accretion rate resulting from the magnetic
stresses will vary with radius, leading to the build-up of material in
some annuli \citep{al01,zh09}, an effect not captured in our
shearing-box calculations.

Apparently the stellar X-rays both increase the disk accretion rate
within 5~AU (figure~\ref{fig:activeht}) and increase the rate at which
material escapes from the disk into a photoevaporating wind at
10-40~AU \citep{de09}.  Furthermore the ionization of the inner disk
could contribute to a wind driven by magnetic activity \citep{si09}.
All these effects tend to reduce the surface density of gas in the
planet-forming region faster around stars with high X-ray
luminosities.  Whether it is possible in this way to account for the
tendency of non-accreting T~Tauri stars to have higher X-ray
luminosities \citep{ns95,sn01,fd03,pk05,tg07} or for an inverse
correlation between X-ray luminosity and stellar accretion rate
\citep{de09} remains unclear.

We suggest that the diversity in the mid-infrared silicate bands among
classical T~Tauri stars is a consequence of the concentration of solid
material associated with planet formation.  The growth of larger
bodies requires a loss of dust, leading generally to a declining
recombination cross-section.  The smallest grains provide most of the
recombination when the particles have a size spectrum appropriate for
the interstellar medium \citep{sm00}, and will dominate still more in
the disk atmosphere, where settling removes the larger grains.  The
disruption of aggregates through collisions (\S\ref{sec:stirring})
thus slows the decline.  However, unless the grains making up the
aggregates also fragment so that the minimum particle size becomes
smaller, the loss of cross-section is likely to continue, since even
1-$\mu$m particles show appreciable settling in the turbulent disk
atmosphere.  The decreasing recombination rates will allow the active
layers to expand into the disk interior, and if the mass column is not
too great, turbulent mixing will eventually reach the midplane,
returning the second-generation debris from planet formation to the
atmosphere.  Questions for future investigation include when and where
the mixing reaches the midplane, and whether the observed correlation
between the degree of settling and the crystalline mass fraction
\citep{wl09} can be explained by the return to the atmosphere of
minerals formed near the midplane at high temperature and pressure in
planetesimal interiors, impacts, or shocks.  Any such correlation will
be blurred by the fact that the degree of dust depletion required for
mixing to reach the midplane depends on the column of gas, the X-ray
luminosity of the star and the magnetic flux delivered to the disk
from the surroundings.

The mid-infrared variability of protostellar disks has been proposed
to arise from the changing illumination of dust in the disk
atmosphere.  Explaining timescales shorter than a week is a challenge,
since the dust in the disks around young Solar-mass stars extends
inward only to about 0.1~AU \citep{mc03,ab05,a08}, where the orbital
period is 12~days.  The results of the MHD calculations demonstrate
that magnetic activity moves dust from place to place in the disk
atmosphere on several timescales.  The optical depth along rays
grazing the surface of the patch of disk frequently varies by a factor
two within a tenth of an orbit as regions of enhanced dust density
pass by.  Over longer time intervals of a few orbits, the photosphere
moves up and down by as much as a scale height due to changes in the
balance between settling and stirring.  Clearly the effects of the
magnetic activity on the dust distribution should be considered when
attempting to understand the infrared variability.  Properly
accounting for the shadows cast on the disk surface requires models of
the inner edge of the dust distribution, perhaps informed by the
time-steady picture that has been developed for Herbig~Ae stars
\citep{dd01,in05}.  Our results suggest that magneto-rotational
turbulence near the dust sublimation radius can cast shadows that vary
over timescales as short as a day.

\acknowledgments

We are grateful for discussions with S. Desch, C. Dullemond,
A. Glassgold, J. Goodman, K. Kretke and M. Wardle.  The work was
carried out in part at the Jet Propulsion Laboratory, California
Institute of Technology with the support of the JPL Research and
Technology Development and NASA Solar Systems Origins Programs.
Copyright 2009.  All rights reserved.


\end{document}